\documentclass[english,aps,prx,superscriptaddress,twocolumn,address,showacknowledgments,longbibliography]{revtex4-1}
\usepackage[latin9]{inputenc}
\usepackage{amsmath}
\usepackage{graphicx}
\usepackage{esint}

\makeatletter
\usepackage{bbm}

\@ifundefined{rangeHsb}{\usepackage{xcolor}}{}

\makeatother

\usepackage{babel}
\begin{document}

\title{Topological quantum fluctuations and travelling wave amplifiers}

\author{Vittorio Peano}

\affiliation{Institute for Theoretical Physics, University of Erlangen-Nürnberg,
Staudtstr. 7, 91058 Erlangen, Germany}

\author{Martin Houde}

\affiliation{Department of Physics, McGill University, 3600 rue University, Montreal,
Quebec, H3A 2T8, Canada}

\author{Florian Marquardt}

\affiliation{Institute for Theoretical Physics, University of Erlangen-Nürnberg,
Staudtstr. 7, 91058 Erlangen, Germany}

\affiliation{Max Planck Institute for the Science of Light, Günther-Scharowsky-Stra{ß}e
1/Bau 24, 91058 Erlangen, Germany}

\author{Aashish A. Clerk}

\affiliation{Department of Physics, McGill University, 3600 rue University, Montreal,
Quebec, H3A 2T8, Canada}

\date{\today}
\begin{abstract}
It is now well-established that photonic systems can exhibit topological
energy bands; similar to their electronic counterparts, this leads
to the formation of chiral edge modes which can be used to transmit
light in a manner that is protected against back-scattering. While
it is understood how classical signals can propagate under these conditions,
it is an outstanding important question how the quantum vacuum fluctuations
of the electromagnetic field get modified in the presence of a topological
band structure. We address this challenge by exploring a setting where
a non-zero topological invariant guarantees the presence of a parametrically-unstable
chiral edge mode in a system with boundaries, even though there are
no bulk-mode instabilities. We show that one can exploit this to realize
a topologically protected, quantum-limited travelling-wave parametric
amplifier. The device is naturally protected both against internal
losses and back-scattering; the latter feature is in stark contrast
to standard travelling wave amplifiers. This adds a new example to
the list of potential quantum devices that profit from topological
transport.
\end{abstract}
\maketitle
The quantization of the electromagnetic field introduces a fundamentally
new phenomenon into physics: vacuum fluctuations that permeate all
of space. These fluctuations were initially seen as a basic unalterable
feature of space-time, before it was realized that they could be engineered
to great effect. Simply modifying geometric boundary conditions changes
the size of the fluctuations as a function of position and frequency,
leading to phenomena such as the Purcell enhancement of spontaneous
emission. The introduction of nonlinear optical materials gives rise
to an even greater level of control, leading to the possibility of
squeezed vacuum states \cite{GerryKnight}, with important applications
to sensing beyond the limits usually set by quantum mechanics \cite{Caves81,Yurke1986,Giovannetti2004}. 

In recent years, new approaches for altering the dynamics of wave
fields have gained prominence, based on engineering periodic materials
to elicit topological properties. Topologically protected unidirectional
wave propagation was originally discovered in the study of 2D electrons
in strong magnetic fields, and underlies the robust quantization of
the Hall conductance \cite{Thouless1982}.  The engineering of topological
photonic materials has been the focus of intense theoretical investigation
\cite{Lu2014}, and various experimental platforms have already been
developed \cite{Wang2009,Kitagawa2012,Rechtsman2013b,Hafezi2013}.
Phononic topological states have also attracted recent attention \cite{Prodan2009,Kane2013,Peano2015,Yang2015,Paulose2015}
and the first experimental steps at the macroscopic scale have been
taken \cite{Susstrunk2015,Nash2015,Paulose2015}. 

Despite this considerable work in topological photonics and phononics,
using topology to address the engineering of vacuum fluctuations has
not been addressed. Most photonic and phononic topological systems
are based on a single particle Hamiltonian which conserves particle
number. These topological states mimic well known electronic topological
phases such as the Quantum Hall phase \cite{Raghu2008a,Raghu2008b,Wang2009,Koch2010,Umucallar2011,Fang2012,Petrescu2012,2014_Lipson_NonreciprocalPhaseShift,Schmidt2015,Peano2015},
or the spin Hall phase \cite{Hafezi2011,Khanikaevphotonic2012,Hafezi2013,Mittal2014,Susstrunk2015}
and have a trivial vacuum. In order to modify the properties of the
vacuum one has to introduce particle non-conserving terms to the Hamiltonian
which can coherently add and remove pairs of particles from the system;
these terms have a formal similarity to pairing terms in the mean-field
description of a fermionic superconductor. If the amplitude of these
terms is sufficiently weak, the system remains stable; even in this
regime, the bosonic nature of the particles makes the topological
properties of such Hamiltonians very different from their fermionic
(topological superconductor) counterparts \cite{Shindou2013,Brandes2015,Bardyn2016,Peano2016}.
An even starker difference occurs when the parametric terms lead to
dynamical instabilities \cite{Barnett2013,Barnett2015}. These instabilities
have no fermionic analogue, and are akin to the parametric instability
in an oscillator whose spring constant is modulated in time.

Here, we consider a situation where parametric driving is introduced
to a system where photons hop on a lattice in the presence of a synthetic
gauge field (see Fig. 1a). We show how to realize an exotic situation
where all bulk modes are stable, but where the topologically-protected
chiral edge modes that exist in a system with a boundary are unstable.
This leads to an unusual spatially-depedent modification of vacuum
fluctuations: when the system is stabilized by dissipation, quantum
fluctuations in the bulk are only weakly perturbed, whereas those
along the system edge are strongly distorted. The result is not just
an unusual driven-dissipative quantum state, but also a unique kind
of photonic device: as we show in detail, the system serves both as
a topologically-protected, non-reciprocal, quantum-limited amplifier,
as well as a source of chiral squeezed light. It thus represents a
potentially powerful new kind of application of topological materials.

\begin{figure*}
\includegraphics[width=2\columnwidth]{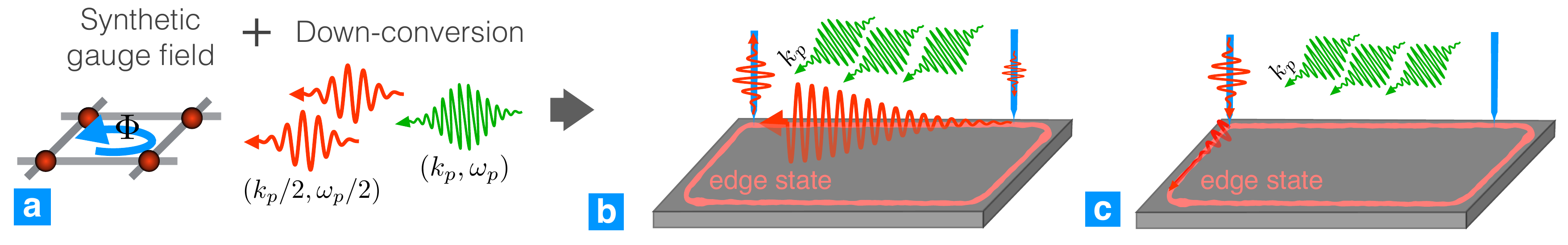}\caption{\label{fig:Set-up-figure}Set up figure: (a) Scheme of the basic interactions:
(i) Photons hopping anti-clockwise around a plaquette pick up a phase
$\Phi$ that can be interpreted as a synthetic gauge field flux. (ii)
Pump photons with frequency $\omega_{p}$ and quasimomentum $k_{p}$
are down-converted into a photon pair with frequency $\omega_{p}$/2
and quasimomentum $k_{p}/2$. (b-c) Combining the two interactions
in a finite geometry allows to engineer a topologically protected
quantum-limited amplifier. A signal injected into the device via a
tapered fiber propagates unidirectionally along the edge. (b) With
the appropriate choice of pump frequency $\omega_{p}$ and quasimomentum
$k_{p}$, the signal is amplified while it travels along the upper
edge. A second tapered fiber detects the amplified signal. (c) When
the input and the output fiber are exchanged the signal propagates
along a different path where it decays due to the lack of phase matching.
This leads to non-reciprocal amplification.}
\end{figure*}

\section{Identifying unstable modes}

Before delving into the details of our proposal, it is useful to discuss
the underlying theoretical ideas in a general setting. Our main goal
is to exploit topological features of a dynamically unstable Hamiltonian,
adding dissipation to realize a non-thermal steady state. In the absence
of topological considerations, this is a situation that is ubiquitous
in quantum optics. The simplest bosonic Hamiltonian exhibiting instability
is the single-mode squeezing Hamiltonian:

\begin{equation}
\hat{H}_{S}=\Delta\hat{a}^{\dagger}\hat{a}+\frac{i}{2}\nu\left(\hat{a}^{\dagger}\hat{a}^{\dagger}-\hat{a}\hat{a}\right),\label{eq:parampHam}
\end{equation}
where $\hat{a}$ is a bosonic annihilation operator. Heuristically,
$\hat{H}_{S}$ describes photons in a single cavity mode (effective
energy $\Delta$) which are subject to coherent two-particle driving
(with amplitude $\nu$). Without dissipation, $\hat{H}_{S}$ becomes
unstable and cannot be diagonalized when the driving amplitude exceeds
the energy cost for creating a pair of photons, i.e. when $\nu>|\Delta|$.
In this regime, the dynamics corresponds to an ever-growing, exponential
accumulation of entangled pairs of bosonic particles: there is no
stationary state. 

If we now add dissipation, stability can be restored by offsetting
the effective two-particle driving described by $\hat{H}_{S}$ against
the decay rate $\kappa$ of the mode; one requires $\kappa>2\sqrt{\nu^{2}-\Delta^{2}}$.
The result is a non-thermal stationary state having a steady flux
of excitations flowing from the driven mode to the dissipative bath
(which could be a waveguide serving as an input-output port). This
is precisely the situation realized in a standard parametric amplifier:
the linear-response properties of this driven-dissipative steady-state
allow for quantum-limited amplification of an additional signal drive.
The required two-photon driving in Eq.(\ref{eq:parampHam}) is generically
realized by using a nonlinearity and parametric down-conversion of
a driven pump mode. 

With these preliminaries, we now consider a very general quadratic
Hamiltonian describing bosons on a lattice subject to parametric driving:
\begin{equation}
\hat{H}=\sum_{\mathbf{k}ss'}\hat{a}_{\mathbf{\mathbf{k},}s}^{\dagger}\mu_{\mathbf{k}ss'}\hat{a}_{\mathbf{k},s'}+\frac{i}{2}\left(\hat{a}_{\mathbf{\mathbf{k},}s}^{\dagger}\nu_{\mathbf{k}ss'}\hat{a}_{\mathbf{k_{p}-k},s'}^{\dagger}-h.c.\right).\label{eq:genericHamiltonian}
\end{equation}
Here the ladder operator $\hat{a}_{\mathbf{k},s}$ annihilates a boson
with quasimomentum $\mathbf{k}$ in the state $s$, where $s,s'=1,\ldots,N$
label polarization and/or sublattice degrees of freedom. The first
set of terms describes the hopping of photons on the lattice, and
explicitly conserves both particle number and quasimomentum. It could
be diagonalized to yield a standard band structure: for each quasimomentum
$\mathbf{k}$, we would have $N$ band eigenstates. The second set
of parametric driving terms break particle number conservation, and
in general also break the conservation of quasimomentum: the two-photon
driving injects pairs with a net quasimomentum $\mathbf{k}_{\mathrm{p}}$,
implying that quasimomentum is only conserved modulo $\mathbf{k}_{\mathrm{p}}$.
For a realization based on a driven $\chi^{(2)}$ medium, the two-photon
driving terms correspond to the down-conversion of pump photons with
quasimomentum $\mathbf{k}_{\mathrm{p}}$ into a pair of photons with
quasimomenta $\mathbf{k}$ and $\mathbf{k}_{\mathrm{p}}-\mathbf{k}$,
respectively. Having a non-zero quasimomentum for injected pairs will
be a crucial resource when we attempt to control parametric instabilities. 

Analogous to the simple Hamiltonian in Eq.(\ref{eq:parampHam}), the
lattice Hamiltonian in Eq. (\ref{eq:genericHamiltonian}) exhibits
instabilities when the amplitude for creating a pair of photons exceeds
the energy of the pair. Formally, one can introduce a generalized
normal mode decomposition of this generic Hamiltonian which explicitly
separates out stable modes and unstable modes. One obtains {[}see
Appendix \ref{sec:Generalized-normal-mode}{]}:
\begin{equation}
\hat{H}=\sum_{\mathbf{k}}\sum_{n\in S_{\mathbf{k}}}E_{n,\mathbf{k}}\hat{n}_{n,\mathbf{k}}+\frac{1}{2}\sum_{n\in U_{\mathbf{k}}}\hat{H}_{n,\mathbf{k}}.\label{eq:normalmodeHamiltonian}
\end{equation}
For each quasimomentum $\mathbf{k}$ in the first Brillouin zone,
we will now have both a set of stable modes (indexed by $n\in S_{\mathbf{k}}$),
and a set of unstable modes ($n\in U_{\mathbf{k}})$. The first $n$-sum
in Eq.(\ref{eq:normalmodeHamiltonian}) describes the stable modes:
they are described by canonical bosonic anihiliation operators $\hat{\beta}_{n,\mathbf{k}}$,
and enter the Hamiltonian in the standard manner, as a real energy
times a number operator $\hat{n}_{n,\mathbf{k}}=\hat{\beta}_{n,\mathbf{k}}^{\dagger}\hat{\beta}_{n,\mathbf{k}}$.
The unstable modes can also be described by canonical bosonic anihiliation
operators $\hat{\beta}_{n,\mathbf{k}}$. They however enter the Hamiltonian
via unstable two-mode squeezing (parametric amplifier) Hamiltonians:

\begin{equation}
\hat{H}_{n,\mathbf{k}}=E_{n,\mathbf{k}}(\hat{n}_{n,\mathbf{k}}-\hat{n}_{n,\mathbf{k_{p}}-\mathbf{k}})+i\lambda_{n,\mathbf{k}}\left(\hat{\beta}_{n,\mathbf{k}}^{\dagger}\hat{\beta}_{n,\mathbf{k_{p}-k}}^{\dagger}-h.c.\right),\label{eq:two-modesqueezing}
\end{equation}
where $E_{n,\mathbf{k}}=-E_{n,\mathbf{k_{p}-k}}$ and $\lambda_{n,\mathbf{k}}=\lambda_{n,\mathbf{k_{p}-}\mathbf{k}}$
are both real. This effective Hamiltonian for the unstable modes has
a simple interpretation: pairs of quasiparticles with opposite energies
$\pm E_{n,\mathbf{k}}$ are created with an amplitude $\lambda_{n,\mathbf{k}}$.
We stress that for any non-zero $\lambda_{n,\mathbf{k}}$, $\hat{H}_{n,\mathbf{k}}$
is unstable (as the total energy for creating the relevant pair of
excitations is always zero). The quasiparticle operators $\hat{\beta}_{n,\mathbf{k}}^{\dagger}$
in Eqs.(\ref{eq:normalmodeHamiltonian},\ref{eq:two-modesqueezing})
are a complete set of Bogoliubov ladder operators.

\begin{figure}
\includegraphics[width=1\columnwidth]{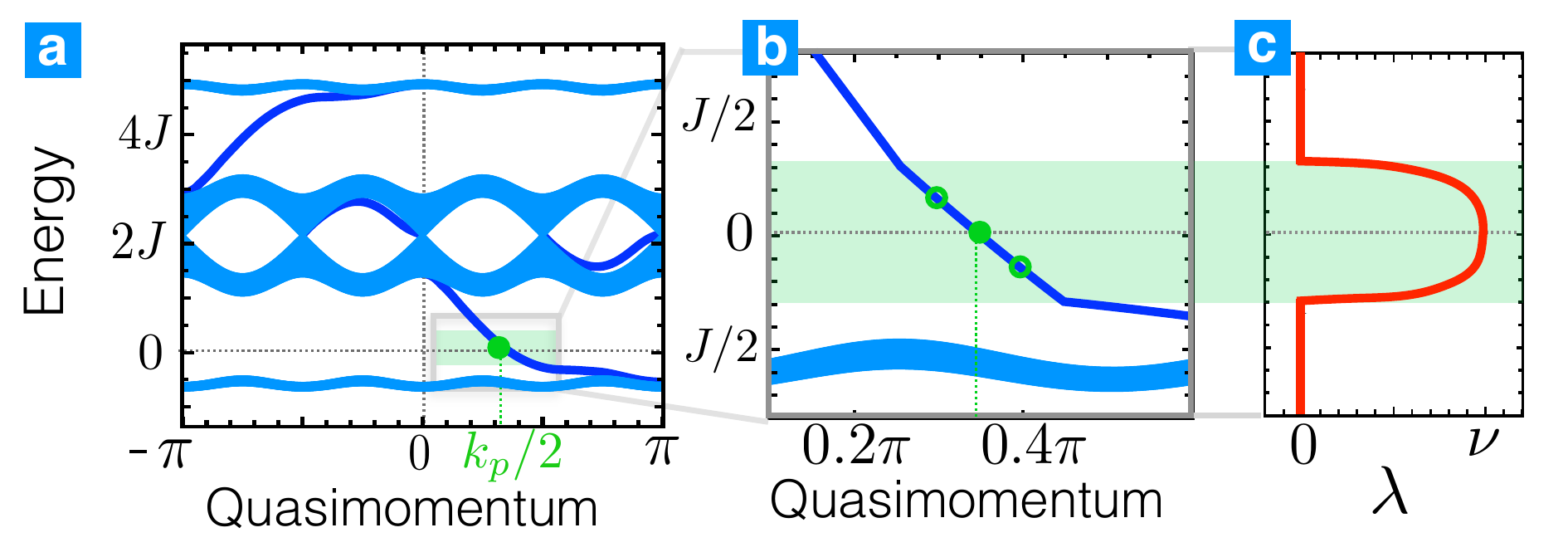}\caption{\label{fig:Band-structure}Topological Band structure. (a) Band structure
of a semi-infinite strip for the Hofstadter model with a flux of $\pi/2$
(when the parametric driving is switched off). The edge states are
plotted in dark blue. The energy is counted off from half of the pump
photon energy. The solid green circle indicates the tuning of the
pump photon quasimomentum $k_{p}$ required to resonantly excite pairs
of down-converted edge state photons with quasimomentum $k_{p}/2$
(at the quasimomentum $k_{p}/2$ the edge state should have energy
$E=0$). (b) Zoom of the band structure for a finite laser power.
In the unstable energy interval highlighted in green, pairs of Bogoliubov
excitations having quasimomenta $k_{p}/2\pm\delta k$ are also excited.
The corresponding amplification amplitude $\lambda$ is shown in panel
(c) {[}Parameters: $\omega_{0}=2.15J$, $\Phi=\pi/2$, $k_{p}=2.2$.
In (b-c) $\nu=0.08J${]} }
\end{figure}

\section{Parametrically driven Hofstadter model}

Having established the necessary theoretical framework, we will now
show how to engineer a Hamiltonian whose only unstable Bogoliubov
modes $\hat{\beta}_{n,\mathbf{k}}^{\dagger}$ are chiral states co-propagating
along the physical boundary of a topological system. We consider photons
hopping on a 2D square lattice in the presence of a synthetic magnetic
field flux, which are also subject to parametric two-photon driving
on each site (see Fig. \ref{fig:Set-up-figure}(c)). Writing the Hamiltonian
in the position basis, we have
\begin{equation}
\hat{H}=\sum_{\mathbf{j}}\omega_{\mathbf{j}}\hat{a}_{\mathbf{j}}^{\dagger}\hat{a}_{\mathbf{j}}-\sum_{\langle\mathbf{i,j}\rangle}J_{\mathbf{ij}}\hat{a}_{\mathbf{i}}^{\dagger}\hat{a}_{\mathbf{j}}+\frac{i}{2}\nu\sum_{\mathbf{j}}\left(e^{i\theta_{\mathbf{j}}}\hat{a}_{\mathbf{j}}^{\dagger}\hat{a}_{\mathbf{j}}^{\dagger}-h.c.\right),\label{eq:Hofstadter}
\end{equation}
where $\hat{a}_{\mathbf{j}}$ is the photon annihilation operator
on site $\mathbf{j}=(j_{x},j_{y})$, and $\omega_{\mathbf{j}}$ are
the corresponding onsite energies. As usual, the synthetic gauge field
is encoded in the pattern of phases $\phi_{\mathbf{ij}}$ of the nearest-neighbor
hopping rates, $J_{\mathbf{ij}}=J\exp(i\phi_{\mathbf{ij}})$. We take
the synthetic flux per plaquette to be $\Phi=\pi/2$. Working in the
Landau gauge, we then have $\phi_{\mathbf{ij}}=0$ for vertical hopping
and $\phi_{\mathbf{ij}}=-\pi j_{y}/2$ for rightwards hopping. The
parametric driving amplitude on a given site $\mathbf{j}$ is written
$\nu e^{i\theta_{\mathbf{j}}}$; we take the phase to vary as $\theta_{\mathbf{j}}=k_{p}j_{x}$,
implying the injection of pairs with a quasimomentum $\mathbf{k}_{\mathrm{p}}=k_{\mathrm{p}}\mathbf{e}_{x}$.
For a realization based on a driven nonlinear medium, $\nu\propto\sqrt{I_{p}}$
where $I_{p}$ is the power of the pump laser, and $k_{p}\mathbf{e}_{x}$
would be the quasimomentum of the pump laser photons. Note that a
gauge transformation $\hat{a}'_{\mathbf{j}}=\hat{a}_{\mathbf{j}}\exp[if_{\mathbf{j}}]$
would modify both pattern of phases $\phi_{\mathbf{ij}}$ and $\theta_{\mathbf{j}}$.

When the laser is switched off, $\nu=0$, and there is no disorder,
$\omega_{\mathbf{j}}=\omega_{0}$, we have the well known Hofstadter
model \cite{Hofstadter1976}. The band structure of a semi-infinite
strip (extending to the lower-half $2D$-plane) is shown in Fig. \ref{fig:Band-structure}a.
The continuous bulk band structure consists of four bands (one for
each site in the magnetic unit cell). The top and bottom bands are
flat Landau levels separated from the two central bands by topological
band gaps. Because of the boundary, one finds inside each topological
band gap an edge state (dark line). The net number of these edge states
(the number weighted by the sign of their slope) is a topologically
protected quantity which does not depend on the shape of the edge
and can be calculated from the bulk Hamiltonian \cite{Hasan2010RMP}.

We now turn on the parametric driving such that the resulting Hamiltonian
can exhibit instability. Our goal is twofold: we want the system to
be unstable \emph{only} if we have a boundary, and in this case, the
dominant unstable Bogoliubov modes should be chiral excitations localized
at the system's boundary. We do this by choosing the parametric drive
parameters so that the only pairs of photons that can be created in
an energy and quasimomentum conserving fashion correspond to edge
state excitations of the original ($\nu=0$) model. For concreteness,
we will focus on exciting the edge mode in the lower topological band
gap (dispersion $\varepsilon_{E}(k)$). In the lab frame, we will
thus tune the pump photon frequency $\omega_{p}$ and quasimomentum
$k_{p}$ so that a single pump photon can be converted into two edge
excitations with frequency $\omega_{p}/2$ and quasimomentum $k_{p}/2$.
In the rotating frame we use to write our Hamiltonians, this requirement
reduces to $\varepsilon_{E}(k_{p}/2)=0$. If this resonance condition
is met, an arbitrarily weak parametric drive $\nu$ will cause instability
of the edge mode. The required tuning is shown in Fig. \ref{fig:Band-structure}a. 

Because of the approximately linear dispersion relation of the edge
mode, the above tuning guarantees that the parametric driving can
resonantly create a pair of edge mode photons having momenta $k_{p}/2\pm\delta k$,
see the hollow circles in Figure \ref{fig:Band-structure}(b). Thus,
even for a weak parametric drive amplitude, the edge state will exhibit
instability over a range of quasimomenta near $k_{p}/2$ (corresponding
to a finite bandwidth around $\omega_{p}/2$ in the lab frame), see
Figure \ref{fig:Band-structure}(c).

Conversely, the energies of two bulk excitations always add up to
a finite value, see Appendix \ref{sec:Appbandstr}. In other words,
all bulk parametric transitions have a finite detuning. This guarantees
the bulk stability (even in the presence of disorder) for a driving
amplitude $\nu$ below the minimal value of the bulk detuning.

\emph{}

\begin{figure}
\includegraphics[width=1\columnwidth]{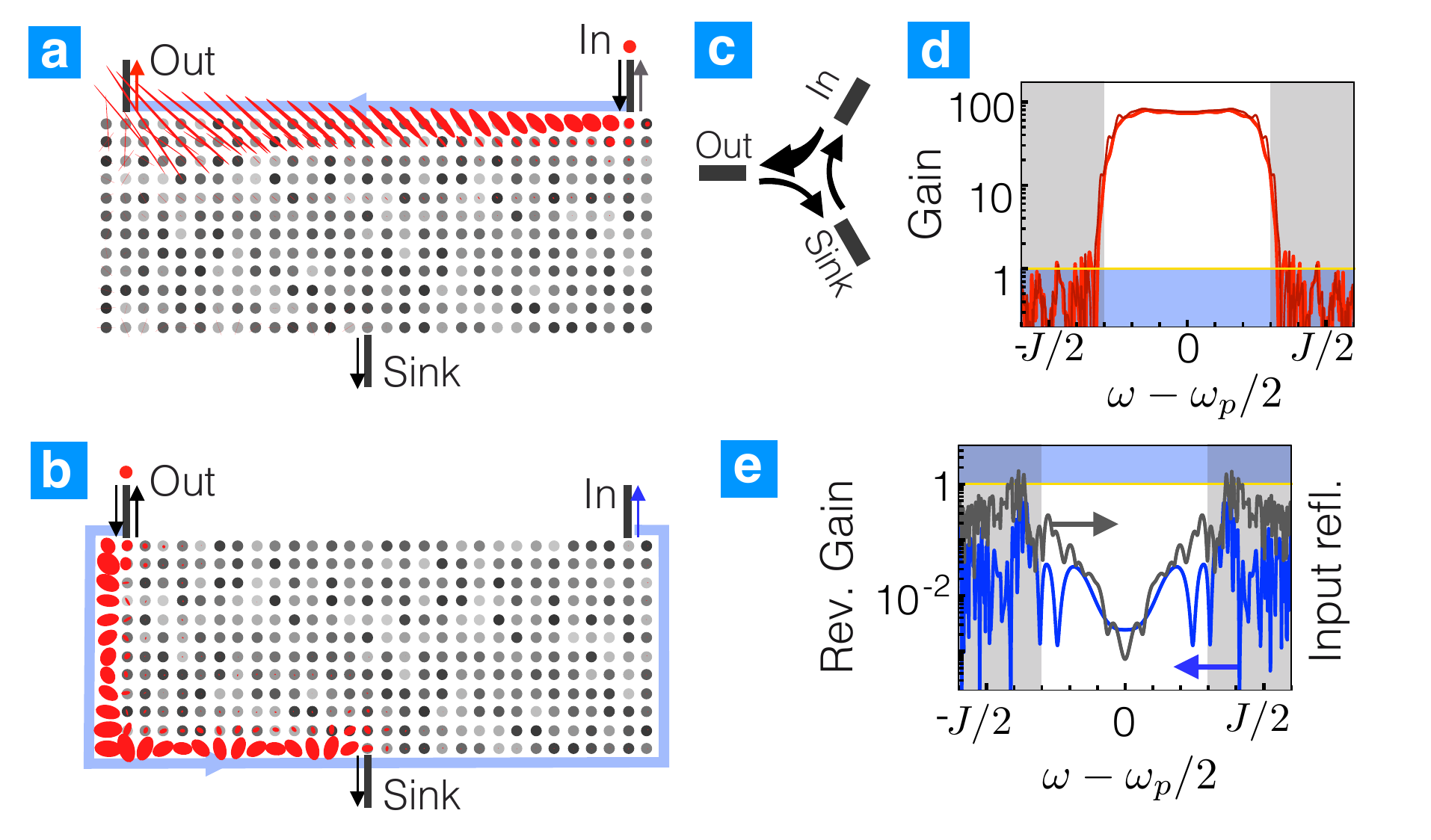}\caption{\label{fig:Linear-response}Linear response of the topological amplifier.
(a-b) The topological amplifier is formed by a $30\times12$ array
of photonic nanocavities. Three of the cavities at the edge of the
sample are attached to waveguides, the input and output ports of the
amplifier and an additional sink port. (c) The amplifier has the geometry
of a circulator. (a-b) The red ellipses represent the linear response
of the field inside the photonic array as a function of the incoming
signal phase. The signal is injected at the port marked by an inward
arrow. (a) A signal injected at the input port propagates unidirectionally
toward the output port. The response is strongly phase sensitive,
a signal with the right phase is amplified along the way. (d) Transmission
power gain for the amplified quadrature as a function of the frequency
of the input signal (counted off from half of the pump frequency)
for a disordered (light thick line) and a clean sample (dark thin
line). The reflection coefficient at the input is shown in panel (e)
for the disordered sample. (b) A signal propagating from the output
port toward the input port follows a different path and it is not
amplified. Moreover, an appropriate matching of the impedances ensures
that it leaks out at the sink port. The resulting (small) reverse
transmission from the output to the input of the amplifier is the
blue curve in panel (e). {[}Parameters: $\omega_{0}=2.14J$, $\Phi=\pi/2$,
$\nu=0.08J$, $k_{p}=2.2$, $\kappa=0.001J$, $\kappa_{{\rm in}}=2.6J$,
$\kappa_{{\rm out}}=3J,$ $\kappa_{{\rm sink}}=4.2J.$ In the disordered
simulations, the offset energies $\delta\omega_{\mathbf{j}}$, represented
by the greyscale in (a-b), are random numbers in the interval $-0.1J<\delta\omega_{\mathbf{j}}<0.1J${]}}
\end{figure}

\section{Topological non-reciprocal amplifier}

Having shown how to realize an unstable topological edge mode, we
now want to understand how one can use it. More precisely, we show
that a finite size array of nanocavities coupled to simple waveguides
can be used as a new kind of topologically-protected, phase-sensitive,
quantum-limited amplifier. The role of the waveguides is two-fold:
they serve as amplifier input-output ports and they stabilize the
dynamics. 

We consider a realization of our system using a $30\times12$ array
of nanocavities, and additionally include three coupling waveguides.
Each waveguide is coupled to a site at the edge of the sample, as
shown in Fig. \ref{fig:Linear-response}a-b. This coupling is described
using standard input/output theory, and is entirely characterized
by the three rates $\kappa_{{\rm in}}$, $\kappa_{{\rm out}},$ and
$\kappa_{{\rm sink}}$, see Appendix \ref{sec:Input-output-formalism}.
In addition, we take each cavity to have an internal-loss decay rate
$\kappa$. 

When the small decay rate $\kappa$ is neglected, and without parametric
driving, the array can be operated as an ideal circulator where a
signal from any waveguide is entirely transmitted into the next waveguide,
see Figure \ref{fig:Linear-response}c. Indeed, it is always possible
to match the impedances at each port to cancel the back-reflection
by tuning the corresponding coupling rate ($\kappa_{{\rm in}}$, $\kappa_{{\rm out}},$
or $\kappa_{{\rm sink}}$). Once inside the array a wave in a topological
band gap has no alternative but to chirally propagate along the edge.
In addition, the impedance matching ensures that a wave impinging
on a waveguide from the edge channel will be entirely transmitted.

We harness the robust non-reciprocity of this topological circulator
to design an amplifier. We use the waveguide on the upper right (left)
as input (output) port of the amplifier. We choose parametric driving
parameters similar as in Fig. \ref{fig:Band-structure}. In the finite
geometry, the quasimomentum matching will be approximately realized
only on the upper edge. Thus, the amplification occurs mainly in the
region between the input and the output port. 

The linear response of the amplifier is investigated numerically in
Fig. \ref{fig:Linear-response}. A signal injected into the array
from the input port propagates chirally until it leaves the array
through the output port, see Fig. \ref{fig:Linear-response}a. Depending
on its phase, it can be amplified or de-amplified along the way. Treating
the amplifier as a phase-sensitive amplifier, we find that the power
gain for the amplified signal quadrature is flat over a large bandwidth,
corresponding to the frequency range over which the edge state dispersion
is purely linear (see panel d). At the same time, any signals incident
upon the output port will be almost entirely dumped into the sink
port, and not reach the input port, see \ref{fig:Linear-response}b.
The residual reverse gain and input reflection are much smaller than
unity, see panel e, ensuring the protection of a potentially fragile
signal source (e.g. a qubit). Crucially, this strongly non-reciprocal
amplifying behavior is of topological origin and is thus robust against
disorder. We demonstrate this resilience by including moderate levels
of disorder in our simulations (see Fig. \ref{fig:Linear-response}).

Our numerical results are in qualitative agreement with analytical
results for a model in which a 1D chiral edge state is coupled to
three waveguides, see Appendix \ref{sec:Appeffectivemodel}. In this
context, we find simple expressions for the impedence matching condition
and the maximum power gain 
\[
\kappa_{i}=\frac{4v}{|u(j_{y}=-1)|^{2}},\quad G\approx\exp\left[\frac{2\nu L}{v}\right],
\]
respectively. Here, $u(j_{y})$ is the transverse edge state wavefunction,
$v$ is the edge state velocity and $L$ is the number of sites separating
the input and the output ports. Thus, we see that the gain is the
exponential of the rate $2\nu$ of creation of down-converted pairs
times the time of flight $L/v$ from the input to the output port.
While we have focused on operation as a phase-sensitive amplifier,
for frequencies different from $\omega_{p}/2$, one could also use
the device as a quantum-limited phase preserving (i.e. non-degenerate)
amplifier.

\subsection*{Quantum-limited amplification }

\begin{figure}
\includegraphics[width=1\columnwidth]{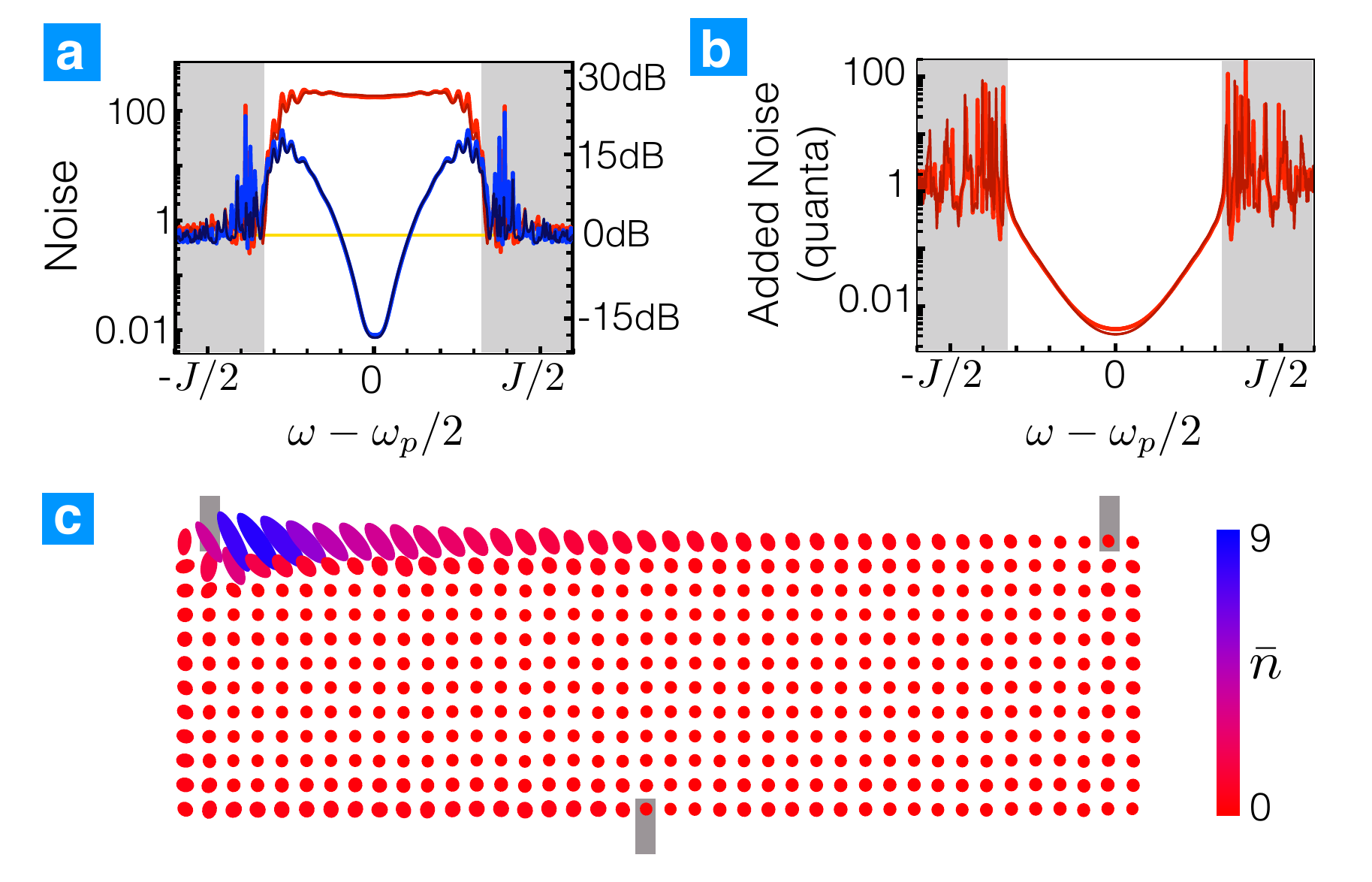}\caption{\label{fig:Quantum-stationary-state}Quantum stationary state and
noise properties of the topological amplifier. (a) Output noise spectra.
The field leaking out of the output waveguide {[}top left corner in
panel (c){]} is strongly squeezed. Plotted are the noise spectral
densities for both the amplified (red) and squeezed (blue) quadratures
of the output field, for both a disordered (light thick lines) and
a clean sample (dark thin lines). The noise in the amplified quadrature
is only slightly larger than the standard quantum limit (i.e. the
amplified vacuum noise from the input port). The excess noise (in
quanta) for both a disordered and a clean sample is shown in panel
(b). (c) Quantum noise ellipses of the field inside the amplifier.
In the bulk, the ellipses have a circular shape and their area is
as small as allowed by the Heisenberg principle, representing a standard
vacuum state. In contrast, the ellipses at the edge are anisotropic
and have areas larger than the shot noise level. This excess noise
does not come from a finite temperature of the environment but rather
by the amplification of the zero-point fluctuations (quantum heating).
Plotted as a color code is also the average number of circulating
photons on each site. The sites attached to waveguides are marked
in grey. ( {[}Parameters: $40\times12$ sites, $\omega_{0}=2.14J$,
$\Phi=\pi/2$, $\nu=0.08J$, $k_{p}=2.2$, $\kappa=0.001J$, $\kappa_{{\rm in}}=2.6J$,
$\kappa_{{\rm out}}=3J,$ $\kappa_{{\rm sink}}=4.2J$. For the disordered
simulations the offset energies $\delta\omega_{\mathbf{j}}$ are random
numbers in the interval $-0.1J<\delta\omega_{\mathbf{j}}<0.1J$. For
all plots, there is only vacuum noise entering from each waveguide.{]} }
\end{figure}

The noise floor of the amplifier is the frequency-resolved noise of
the amplified output quadrature, see Appendix \ref{sec:Input-output-formalism}.
It is plotted in dark (light) red for a clean (disordered) sample
in Fig. \ref{fig:Quantum-stationary-state}a. The quantum limit on
a phase-sensitive amplifier is to have zero added noise, implying
that the noise floor is simply set by the amplification of the vacuum
fluctuations entering the input port. The added noise (expressed as
an equivalent number of input quanta) is plotted in Fig. \ref{fig:Quantum-stationary-state}b;
despite disorder and noise associated with internal loss, the amplifier
is nearly quantum limited over the entire amplification bandwidth. 

\section{Topological stationary state\emph{ }}

Next, we investigate the quantum fluctuations in the stationary state
that arise from the steady flow of photons from the parametric pump
to the amplifier ports in the form of down-converted radiation. Physically,
such a flow arises because vacuum fluctuations entering the input
port (within the amplification bandwidth) are amplified inside the
array before exiting through the output port. The resulting stationary
state of each cavity is Gaussian, and can be represented by a Wigner
function; these are visualized as a set of noise ellipses in Fig.
\ref{fig:Quantum-stationary-state}c. Each noise ellipse completely
characterizes the steady state of the corresponding site once all
remaining sites and the waveguides have been traced out. The areas
of the ellipses are constrained from below by the Heisenberg uncertainty
principle and assume the minimal possible value for pure states. The
bulk sites are all in the trivial quantum groundstate, which is characterized
by circular ellipses with the minimal area and zero photons. On the
other hand, the eccentricity, the area and the average on-site photon
number (color code) grow while moving from the input to the output
port along the upper edge (the major axis corresponds to the amplified
quadrature). We emphasize that in a thermal equilibrium setting, the
area of the ellipses would be equal on all sites and directly reflect
the environment temperature. Here, the excess noise of the sites on
the upper edge has a quantum origin. This phenomenom has been termed
quantum heating \cite{Dykman2011,Peano2010,Lemonde2015}. 

Due to quantum heating, the stationary state of each cavity along
the edge corresponds to a thermal squeezed state, implying that the
squeezed quadrature has a larger variance than required by the uncertainty
principle. Nonetheless, the frequency-resolved output noise is strongly
squeezed below the vacuum level (for frequencies within the amplification
bandwidth). Remarkably, the quality of the output squeezing is not
deteriorated in the presence of disorder. For the parameters considered
here, more than $15{\rm dB}$ of output squeezing are predicted both
in the case of a clean and a disordered sample, see the blue curves
in Fig. \ref{fig:Quantum-stationary-state}a.

\section{Implementation }

Photonic gauge fields have been already realized in several experimental
platforms \cite{Wang2009,Rechtsman2013b,Hafezi2013}. The only additional
ingredient of our proposal is the parametric pumping with a finite
quasimomentum. In the setup of Ref. \cite{Hafezi2013}, one could
fabricate the microrings from a nonlinear optical $\chi^{(2)}$ material
and drive them with a single laser impinging at a finite angle. The
implementation of parametric pumping \cite{Peano2016} and synthetic
gauge fields \cite{Koch2010,Fang2012,Schmidt2015,Peano2015,Peano2016}
is in principle possible in any cavity array platform where a nonlinear
resource is available. These include photonic crystals microcavities
\cite{Notomi2008} fabricated from nonlinear optical $\chi^{(2)}$
materials \cite{Yariv2002,Eggleton2011,Dahdah2011} or piezoelectric
materials, optomechanical arrays based on optomechanical crystals
\cite{Painter2013,Regal2013}, or lattices of superconducting resonators
with embedded Josephson nonlinearities \cite{Houck2012}. We also
note that very recently, a proposal for realizing topological insulator
physics in a classical optical network with nonlinearities was put
forward \cite{Cirac2016}; such a setup could also be adapted to implement
our scheme, as it contains all the necessary ingredients.

\section{Conclusions and outlook}

In this work, we have introduced a means to tie the squeezing and
amplification of vacuum fluctuations to topological properties of
a band structure. Our work represents a new design principle for a
non-reciprocal quantum-limited amplifier which has topological protection.
Non-reciprocal amplifiers have the potential to revolutionize experiments
with superconducting qubits, as they could eliminate the need for
ferrite-based circulators and the accompanying insertion losses which
limit current experiments. A variety of (non-topological) designs
based on multiple parametric interactions have been proposed recently
\cite{Metelmann2015,Ranzani2015} and even realized experimentally
\cite{Abdo2014,Sliwa2015}, including a travelling wave parametric
amplifier (TWPA) built using an array of over 2000 Josephson junctions
\cite{Macklin2015}. In such a conventional TWPA, the reverse transmission
is at best unity, and even small amounts of disorder can lead to large
amounts of unwanted reflection gain. In contrast, the topological
underpinnings of our design ensure reverse transmission and input
reflection coefficients that are well below unity even in the presence
of disorder. 

More generally, our topological amplifier differs markedly from other
proposed topological devices such as isolators or non-amplifying circulators
\cite{Lu2014}, in that it has some protection against internal losses:
in the large gain limit, only the loss (and corresponding noise) in
the immediate vicinity of the input port hinder quantum limited operations,
as it is only this noise which is amplified to any significant degree
(see Appendix \ref{sec:Applosses} for a quantitative discussion of
this point).

In conclusion, our work shows how utilizing the topological properties
of an unstable bosonic Hamiltonian provides a new route for both engineering
electromagnetic vacuum fluctuations, and correspondingly, constructing
a new kind of topologically-protected, non-reciprocal quantum amplifier.
It opens the door to future studies, both pursuing other kinds of
novel applications, as well as more fundamental issues (e.g. the effects
of additional photon-photon interactions in such systems).

\section*{Acknowledgements}

V.P., C.B., and F.M. acknowledge support by an ERC Starting Grant
OPTOMECH, by the DARPA project ORCHID, and by the European Marie-Curie
ITN network cQOM. M.H. and A.A.C. acknowledge support from NSERC.
We thank Ignacio Cirac, Sebastian Huber, and André Xuereb for discussion.

\appendix

\section{Generalized normal mode decomposition\label{sec:Generalized-normal-mode}\emph{ }}

We consider the generic Hamiltonian Eq. (\ref{eq:genericHamiltonian}).
We group all ladder operators with quasimomentum $\mathbf{k}$ in
a vector of ladder operators, $|\hat{a}_{\mathbf{k}}\rangle=(\hat{a}_{\mathbf{k},1},\ldots,\hat{a}_{\mathbf{k},N},\hat{a}_{\mathbf{k_{p}-k},1}^{\dagger},\ldots,\hat{a}_{\mathbf{k_{p}-k},N}^{\dagger})^{T}$.
 The Heisenberg equation of motion for $|\hat{a}_{\mathbf{k}}\rangle$
reads
\begin{equation}
\frac{d}{dt}|\hat{a}_{\mathbf{k}}\rangle=-i\sigma_{{\rm z}}h_{\mathbf{k}}|\hat{a}_{\mathbf{k}}\rangle,\label{eq:Heisenberg}
\end{equation}
where 
\[
\sigma_{{\rm z}}=\begin{pmatrix}1\!\!1_{N} & 0\\
0 & -1\!\!1_{N}
\end{pmatrix}
\]
and $h_{\mathbf{k}}$ is the Bogoliubov de Gennes Hamiltonian 
\[
h_{\mathbf{k}}=\begin{pmatrix}\mu_{\mathbf{k}} & i\nu_{\mathbf{k}}\\
-i\nu_{\mathbf{k}}^{\dagger} & \mu_{\mathbf{k_{p}-k}}
\end{pmatrix}.
\]

In the following, we explicitly construct a complete set of Bogoliubov
operators $\hat{\beta}_{n,\mathbf{k}}$, leading to the generalized
normal mode decomposition Eq. (\ref{eq:normalmodeHamiltonian}), from
the solutions of the eigenvalue problem
\begin{equation}
\sigma_{{\rm z}}h_{\mathbf{k}}|\mathbf{k}_{n,l}\rangle=\Lambda_{\mathbf{k},n,l}|\mathbf{k}_{n,l}\rangle.\label{eq:eigenequation}
\end{equation}
We note that we need to find only $N$ annihilation operators while
the eigenvalue problem has dimension $2N$. However, the equations
for quasimomentum $\mathbf{k_{p}-k}$ are not independent from the
equations for quasimomentum $\mathbf{k}$. The ones can be obtained
from the others by taking the adjoint. This doubling of the degrees
of freedom accompanied by an embedded particle-hole symmetry occurs
because we are effectively applying a single-particle formalism to
a problem where the number of excitations is not conserved. 

From Eq. (\ref{eq:eigenequation}) it is easy to prove that eigenvalues
which are not related by complex conjugation $\Lambda_{n,l,\mathbf{k}}\neq\Lambda_{n',l',\mathbf{k}}^{*}$
have $\sigma_{z}$-orthogonal eigenvectors, $\langle\mathbf{k}_{n,l}|\sigma_{{\rm z}}|\mathbf{k}_{n',l'}\rangle=0$.
Moreover, the eigenvectors with real eigenvalues have a non-zero symplectic
norm, $\langle\mathbf{k}_{n,l}|\sigma_{{\rm z}}|\mathbf{k}_{n,l}\rangle\neq0$
(which can also be negative). We assign the label $l=+$ to the positive
norm eigenvectors. We construct a set of orthonormal Bogoliubov creation
operators from these positive norm solutions with the definition,
\begin{equation}
\hat{\beta}_{n,\mathbf{k}}\equiv\langle\mathbf{k}_{n,+}|\sigma_{z}|\hat{a}_{\mathbf{k}}\rangle.\label{eq:Bogtrafostable}
\end{equation}
We note that the scalar product between a standard vector and a vector
of operators is an operator. Moreover, we have to normalize the positive
vectors $|\mathbf{k}_{n,l}\rangle$ according to $\langle\mathbf{k}_{n,+}|\sigma_{{\rm z}}|\mathbf{k}_{n,+}\rangle=1$
such that $[\hat{\beta}_{n,\mathbf{k}},\hat{\beta}_{n',\mathbf{k}}^{\dagger}]=\delta_{nn'}.$
By taking the time derivative of Eq. (\ref{eq:Bogtrafostable}) and
plugging Eq. (\ref{eq:Heisenberg}) and the adjoint of Eq. (\ref{eq:eigenequation})
we immediately find
\[
\dot{\hat{\beta}}_{n,\mathbf{k}}=-i\langle\mathbf{k}_{n,+}|\sigma_{z}\sigma_{z}h_{\mathbf{k}}|\hat{a}_{\mathbf{k}}\rangle=-i\Lambda_{n,+,\mathbf{k}}\hat{\beta}_{n}
\]
Thus, $\hat{\beta}_{n,\mathbf{k}}$ is the annihilation operator of
a harmonic oscillator with energy $E_{n,\mathbf{k}}=\Lambda_{n,+,\mathbf{k}}$.
In the same way, one could construct a set of creation operators $\hat{\beta}_{n,\mathbf{k_{p}-k}}^{\dagger}$
from the negative norm eigenvectors $|\mathbf{k}_{n,-}\rangle$. However,
it is possible to focus only on the positive norm solutions because
of the particle-hole symmetry: the information encoded in the negative
norm solutions $|\mathbf{k}_{n,-}\rangle$ is also encoded in the
positive norm solutions $|(\mathbf{k_{p}-k})_{n,+}\rangle$. 

The remaining eigenvectors have zero norm, $\langle\mathbf{k}_{n,\pm}|\sigma_{{\rm z}}|\mathbf{k}_{n,\pm}\rangle=0$.
They appear whenever the Hamiltonian is unstable. In this case, the
matrix $\sigma_{z}h_{\mathbf{k}}$ has pairs of complex conjugated
eigenvalues $\Lambda_{n,+,\mathbf{k}}=\Lambda_{n,-,\mathbf{k}}^{*}.$
For concreteness, we indicate with the label $+$ the eigenvalues
with positive imaginary part. The pair of eigenvectors $|\mathbf{k}_{n,\pm}\rangle$
are not ortoghonal to each other, $\langle\mathbf{k}_{n,+}|\sigma_{{\rm z}}|\mathbf{k}_{n,-}\rangle\neq0$.
In this case, we define a pair of commuting ladder operators as,
\begin{eqnarray}
\hat{\beta}_{n,\mathbf{k}} & \equiv & \frac{1}{\sqrt{2}}\left(\langle\mathbf{k}_{n,-}|+i\langle\mathbf{k}_{n,+}|\right)\sigma_{z}|\hat{a}_{\mathbf{k}}\rangle,\label{eq:bogtrafounst}\\
\hat{\beta}_{n,\mathbf{k_{p}-k}}^{\dagger} & \equiv & \frac{1}{\sqrt{2}}\left(\langle\mathbf{k}_{n,-}|-i\langle\mathbf{k}_{n,+}|\right)\sigma_{z}|\hat{a}_{\mathbf{k}}\rangle.\label{eq:bogtrafousdag}
\end{eqnarray}
The bosonic commutation relations are recovered by requiring the normalization
$\langle\mathbf{k}_{n,-}|\sigma_{{\rm z}}|\mathbf{k}_{n,+}\rangle=i.$
By taking the time derivative of Eq. (\ref{eq:bogtrafounst}) and
using Eq. (\ref{eq:Heisenberg}), the transpose of Eq. (\ref{eq:eigenequation}),
and Eq. (\ref{eq:bogtrafousdag}) we find
\[
\dot{\hat{\beta}}_{n,\mathbf{k}}=-i{\rm Re}[\Lambda_{n,+,\mathbf{k}}]\hat{\beta}_{n,\mathbf{k}}+{\rm Im}[\Lambda_{n,+,\mathbf{k}}]\hat{\beta}_{n,\mathbf{k_{p}-k}}^{\dagger}.
\]
Likewise, we find
\[
\dot{\hat{\beta}}_{n,\mathbf{k_{p}-k}}^{\dagger}=-i{\rm Re}[\Lambda_{n,+,\mathbf{k}}]\hat{\beta}_{n,\mathbf{k_{p}-k}}^{\dagger}+{\rm Im}[\Lambda_{n,+,\mathbf{k}}]\hat{\beta}_{n,\mathbf{k_{p}-k}}.
\]
The corresponding Hamiltonian is the two-mode squeezing Hamiltonian
(Eq.(\ref{eq:two-modesqueezing})) with energies $E_{n,\mathbf{k}}=-E_{n,\mathbf{k_{p}-k}}={\rm Re}[\Lambda_{n,+,\mathbf{k}}]$
and amplification amplitude $\lambda_{n,\mathbf{k}}={\rm Im}[\Lambda_{n,+,\mathbf{k}}].$ 

When the matrix $\sigma_{z}h$ is diagonalizable, the set of Bogoliubov
annihilation operators defined in Eqs. (\ref{eq:Bogtrafostable})
and (\ref{eq:bogtrafounst}) is complete. The pathological case where
the matrix $\sigma_{z}h$ is not diagonalizable occurs only exactly
at the threshold of an instability.

\section{Details of the calculation of the band structure\label{sec:Appbandstr}}

\subsection*{Stability of the bulk Hamiltonian}

\begin{figure*}[h]
\centering{}\includegraphics[width=15cm]{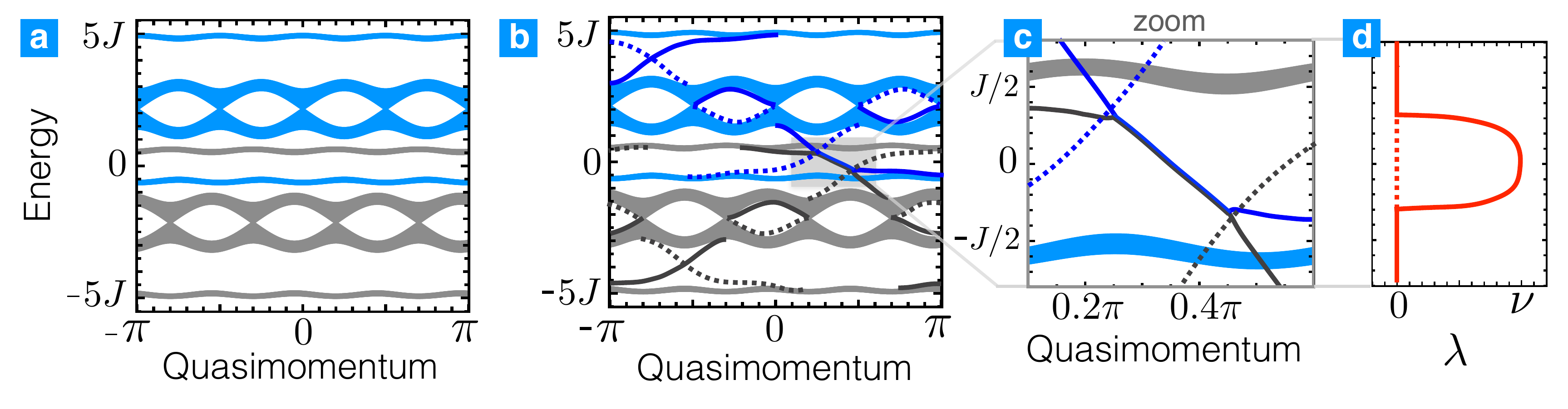}
\caption{ \label{FigSup1} Bogoliubov de Gennes band structures. (a) BdG band
structure for a system without boundaries. The stability of the bulk
is protected by the complete band gaps separating neighboring particle
and hole bands. (b-c) BdG band structure of a finite strip of width
$M=40$ magnetic unit cells. (d) Corresponding amplification amplitudes
$\lambda_{n}$. The parameters are the same as for Figure 2 of the
main text {[}$\omega_{0}=2.15J$, $\Phi=\pi/2$, $k_{p}=2.2$, $\nu=0.08J$.{]}}
\end{figure*}

In the main text, we have explained that a bosonic Hamiltonian with
anomalous pairing terms is unstable when it allows the creation of
a pair of Bogoliubov excitations without any net energy change. In
other words, the sum of two quasiparticle energies should be zero.
We have also claimed that for the parameters of Figure 2 the bulk
Hamiltonian is stable because no combination of bulk states whose
energies add up to zero exists. This is not immediately obvious from
the plot of the standard band structure. In order to visually illustrated
the absence of such combination of bulk states one has rather to plot
the corresponding Bogolibov de Gennes band structure, see Fig.~\ref{FigSup1}.
For each value of $k_{x}$, the latter is formed by both the quasiparticle
energies $E_{n,k_{x}}$ (blue bands) as well as the energies $-E_{n,k_{p}-k_{x}}$
(grey bands). In analogy to the descriptions of quasiparticles in
fermionic superconductors, one refers to the former (latter) as the
particle (the hole) bands of the system. The absence of crossings
between particle and hole bands for the chosen pump laser frequency
implies that there is no combination of a pair of bulk quasiparticles
whose energies add up to zero while at the same time the corresponding
quasimomenta add up to $k_{p}$. In the limit of vanishing parametric
driving ($\nu=0$), the minimal distance between particle and hole
bands is the minimal detuning of a parametric transition involving
two bulk states.

In the presence of weak disorder, there is no selection rule for the
quasimomenta of the pair of quasiparticles created in a parametric
transition. Nevertheless, all possible parametric transitions are
still detuned because the band gap separating particle and hole bands
is not merely locally defined (for a fixed $k_{x}$) but rather extends
to the whole Brillouin zone. Thus, the stability of the bulk Hamiltonian
is a robust feature.

As explained in Appendix \ref{sec:Generalized-normal-mode}, the Bogoliubov
de Gennes band structure can be calculated by diagonalizing the matrix
$\sigma_{z}h(k_{x})$ where $h(k_{x})$ is the first-quantized Bogoliubov
de Gennes Hamiltonian equivalent to the second quantized Hamiltonian
Eq.~(5) of the main text in the presence of periodic boundary conditions
both in the $x$- and $y$- directions.

\subsection*{Band structure of a strip}

Figure 2 of the main text represents the band structure and the amplification
amplitudes of a semi-infinite strip. Those have been derived from
the the Bogoliubov de Gennes band structure and the amplification
rates of a strip with two physical edges as explained below.

We have simulated a finite strip of width $M=40$ magnetic unit cells.
For each value of the quasimomentum $k_{x}$, the set of energies
$E_{n,k_{x}}$ and $-E_{n,k_{p}-k_{x}}$ forming the BdG band structure
and the corresponding amplification rates $\lambda_{n}(k_{x})$ are
calculated by diagonalizing the relevant $4M\times4M$ matrix $\sigma_{z}h(k_{x})$.
While strictly speaking the energy spectrum is discrete, the spacing
between subsequent bulk states is not visible on the figure scale.
The resulting band structure {[}panels (b-c){]} and the corresponding
amplification rates {[}panel (d){]} are shown in Fig.~\ref{FigSup1}.
By inspecting the corresponding wavefunctions one can easily distiguish
between particle and hole bands (plotted in blue and grey, respectively),
edge and bulk states (plotted by the thick dark and thin light lines,
respectively) and upper- and lower-edge states (solid and dashed lines,
respectively).

Since the hole bands are not independent from the particle bands and
refer to the same set of normal modes, we have not displayed the hole
energies in the plots in the main text. Moreover, we have shown only
the edge states localized on the upper edge. Removing the edge states
localized on the lower edge effectively corresponds to plotting the
band structure of a semi-infinite strip extending to the lower half
plane.

We note that the particle (hole) edge states cross the hole (particle)
band. Thus, a pair formed by an edge and a bulk excitation can in
principle also be resonantly excited. However, the corresponding matrix
element is very small and the resulting amplification rate is not
visible on the scale of the figure.

\subsection*{Analytical derivation of Edge state dispersion and amplification
rate}

Here, we outline a direct derivation of the edge state dispersion
for a semi-infinite strip.

We first, consider the case where the parametric driving is switched
off ($\nu=0$). The Hamiltonian in terms of the annihilation operators
$\hat{a}_{j_{y},k_{x}}$ of a plane wave with quasimomentum $k_{x}$
on the $|j_{y}|$-th row reads
\begin{eqnarray}
\hat{H} & = & \sum_{j_{y},k_{x}}\left(\omega_{0}-2J\cos(k_{x}+\pi j_{y}/2)\right)\hat{a}_{j_{y},k_{x}}^{\dagger}\hat{a}_{j_{y},k_{x}}\nonumber \\
 &  & -J\left(\hat{a}_{j_{y}-1,k_{x}}^{\dagger}\hat{a}_{j_{y},k_{x}}+\hat{a}_{j_{y},k}^{\dagger}\hat{a}_{j_{y}-1,k_{x}}\right).\label{eq:Hofstadter_strip}
\end{eqnarray}
 Since, the strip extends to the lower-half plane the $j_{y}$-th
sum runs over the negative integers. By definition of the normal modes
$\hat{\alpha}_{n,k_{x}}$ and eigenenergies $E_{n}[k_{x}]$ we have
$\hat{H}=\sum_{n,k_{x}}E_{n}[k_{x}]\hat{\alpha}_{n,k_{x}}^{\dagger}\hat{\alpha}_{n,k_{x}}$,
or equivalently $[\hat{\alpha}_{n,k_{x}},\hat{H}]~=~E_{n}[k_{x}]\hat{\alpha}_{n,k_{x}}$.
By plugging the ansatz $\hat{\alpha}_{n,k_{x}}^{\dagger}=\sum_{j_{y}}u_{n,k_{x}}[j_{y}]\hat{a}_{j_{y},k_{x}}^{\dagger}$,
we arrive to the first-quantized time-independent Schrödinger equation
which we set in the matrix form, 
\begin{align}
 & \left(\begin{array}{c}
u_{n,k_{x}}[j_{y}-1]\\
u_{n,k_{x}}[j_{y}]
\end{array}\right)=\tilde{M}_{j_{y}}(\epsilon_{n}[k_{x}])\left(\begin{array}{c}
u_{n,k_{x}}[j_{y}]\\
u_{n,k_{x}}[j_{y}+1]
\end{array}\right)\nonumber \\
 & =\left(\begin{array}{cc}
-\epsilon_{n}[k_{x}]-2\cos(k_{x}+\pi j_{y}/2) & -1\\
1 & 0
\end{array}\right)\left(\begin{array}{c}
u_{n,k_{x}}[j_{y}]\\
u_{n,k_{x}}[j_{y}+1]
\end{array}\right).
\end{align}
Here, we have defined $\epsilon_{n}[k_{x}]=(E_{n}[k_{x}]-\omega_{0})/J$.
In principle, the above equation holds only for $j_{y}\leq-2$ but
not for $j_{y}=-1$ because there is no row corresponding to $j_{y}=0$.
One can circumvent this problem by formally adding the row $j_{y}=0$
together with the boundary condition $u_{n,k_{x}}[0]=0$. Thus, one
immediately finds 
\begin{align}
\left(\begin{array}{c}
u_{n,k_{x}}[j_{y}]\\
u_{n,k_{x}}[j_{y}-1]
\end{array}\right) & =\prod_{|j'_{y}|<|j_{y}|}\tilde{M}_{j'_{y}}(\epsilon_{n}[k_{x}])\left(\begin{array}{c}
u_{n,k_{x}}[-1]\\
0
\end{array}\right).\label{eq:matrixeom1}
\end{align}
Next, we focus on the edge state solutions. In the following, we drop
the subscript $n$ because there is only one edge state in each band
gap, see below. For the edge states we have to enforce the boundary
condition 
\begin{equation}
\lim_{j_{y}\to-\infty}u_{k_{x}}[j_{y}]=0
\end{equation}
Equivalently, one can require
\begin{eqnarray}
 &  & \lim_{m\to\infty}\left(\begin{array}{c}
u_{k_{x}}[-4m-1]\\
u_{k_{x}}[-4m]
\end{array}\right)=\nonumber \\
 &  & \lim_{m\to\infty}\left(M(\epsilon[k_{x}])\right)^{m}\left(\begin{array}{c}
u_{k_{x}}[-1]\\
0
\end{array}\right)=0.\label{eq:matrixeom2}
\end{eqnarray}
 where $M(\epsilon[k_{x}])$ is the transfer matrix by a full magnetic
unit cell (four sites), 
\begin{equation}
M(\epsilon[k_{x}])=\tilde{M}_{-4}(\epsilon[k_{x}])\tilde{M}_{-3}(\epsilon[k_{x}])\tilde{M}_{-2}(\epsilon[k_{x}])\tilde{M}_{-1}(\epsilon[k_{x}]).
\end{equation}
(5) is fullfilled if and only if the vector $(1,0)$ is an eigenvector
of the $4$-site transfer matrix $M(\epsilon[k_{x}])$ whose eigenvalue
has modolous smaller than unity. In other words, the edge state dispersion
is determined by the conditions,
\begin{eqnarray}
M_{2,1}(\epsilon[k_{x}]) & = & -\epsilon(-4+\epsilon^{2}+2\cos[2k_{x}])\nonumber \\
 &  & +2\cos[k_{x}](\epsilon^{2}-4\sin[k_{x}]^{2})=0,\label{eq:polenergy}\\
|M_{1,1}(\epsilon[k_{x}])| & = & |\epsilon^{4}-7\epsilon^{2}-2\epsilon(\cos[k_{x}]+\sin[k_{x}])\nonumber \\
 &  & +3+2\sin[2k_{x}]-4\cos[4k_{x}]|<1.\nonumber 
\end{eqnarray}
 Equation (6) is a third order polynomial in the dimensionless energy
$\epsilon$. Thus, it has three roots for each value of $k_{x}$.
Each root corresponds to a solution inside one of the three band gaps
(lower, middle, or upper). By plugging the analytical expression of
each root of Equation (6) into Equation (7) we find the range of $k_{x}$
where the corresponding edge state is defined: the interval $-\pi<k_{x}<0$
for the upper edge state, $0<k_{x}<\pi$ for the lower edge state,
and the two intervals $-\pi/2<k_{x}<0$ and $\pi/2<k_{x}<\pi$ for
the edge state in the local band gap between the two central bands.
In order to enforce the bosonic commutation relations $[\hat{\alpha}_{k_{x}},\hat{\alpha}_{k_{x}}^{\dagger}]=1$,
we have to appropriately normalize the single-particle wavefunctions
$\sum_{j_{y}}|u_{k_{x}}[j_{y}]|^{2}=1$. This condition fixes the
modolous of the wavefunction on the initial site $u_{k_{x}}[-1]$.
The complex phase of $u_{k_{x}}[-1]$ is arbitrary. For concreteness,
we choose $u_{k_{x}}[-1]$ to be real.

Next, we show how the edge state is modified by the parametric pump.
The pump drive written in terms of the annihilation operators $\hat{a}_{j_{y},k_{x}}$
reads, 
\begin{equation}
\hat{H}_{{\rm pump}}=i\frac{\nu}{2}\sum_{j_{y},k_{x}}\left(\hat{a}_{j_{y},k_{x}}^{\dagger}\hat{a}_{j_{y},k_{p}-k_{x}}^{\dagger}-\hat{a}_{j_{y},k_{x}}\hat{a}_{j_{y},k_{p}-k_{x}}\right).
\end{equation}
Before pursuing an exact numerical solution we adopt a semi-analytical
treatment. We focus on the lower band gap edge state which is resonantly
driven for the parameters of Fig. 2 of the main text. As in the main
text, we consider the case where the frequency and quasimomentum of
the pump laser are chosen to resonantly excite pair of edge photons
with quasimomentum $k_{p}/2$, 
\begin{equation}
\omega_{0}=-J\varepsilon[k_{p}/2].
\end{equation}
In this case, it is convenient to introduce the quasimomentum $\delta k=k_{x}-k_{p}/2$
counted off from $k_{p}/2$. By rewriting the full Hamiltonian (including
the pump) in terms of the edge states ladder operators $\hat{\alpha}_{\delta k}$
and keeping only the terms that acts on the lower edge state (we neglect
an off-resonant parametric coupling to the bulk modes and the other
edge states), we find $H^{{\rm (edge)}}=\sum_{\delta k}H_{\delta k}^{{\rm (edge)}}/2$
\begin{eqnarray}
H_{\delta k}^{{\rm (edge)}} & = & \tilde{E}_{n}(\delta k)(\hat{\alpha}_{\delta k}^{\dagger}\hat{\alpha}_{\delta k}-\hat{\alpha}_{-\delta k}^{\dagger}\hat{\alpha}_{-\delta k})\nonumber \\
 &  & +\Delta[\delta k](\hat{\alpha}_{\delta k}^{\dagger}\hat{\alpha}_{\delta k}+\hat{\alpha}_{-\delta k}^{\dagger}\hat{\alpha}_{-\delta k})\nonumber \\
 &  & +iV[\delta k]\left(\hat{\alpha}_{\delta k}^{\dagger}\hat{\alpha}_{-\delta k}^{\dagger}-\hat{\alpha}_{\delta k}\hat{\alpha}_{-\delta k}\right).\label{eq:Hamedge}
\end{eqnarray}
 Here, we have grouped all excitation conserving terms into two contributions
whose amplitudes $\tilde{E}_{n}(\delta k)$ and $\Delta_{n}(\delta k)$
are an odd an even function of $\delta k$, repectively, 
\begin{eqnarray}
\tilde{E}_{n}(\delta k) & = & J\sum_{n\ge1}\frac{{\rm d}^{2n-1}\varepsilon}{{\rm d}k_{x}^{2n-1}}\Big|_{k_{p}/2}\frac{\delta k^{2n-1}}{(2n-1)!}\label{eq:energyedge}\\
\Delta(\delta k) & = & J\sum_{n\ge1}\frac{{\rm d}^{2n}\varepsilon}{{\rm d}k_{x}^{2n}}\Big|_{k_{p}/2}\frac{\delta k^{2n}}{(2n)!}.\label{eq:detedge}
\end{eqnarray}
Note that $\Delta(\delta k)$ is the quasimomentum dependent detuning
of the parametric transition creating pairs of photons with quasimomentum
$k_{p}/2\pm\delta k$. The corresponding parametric coupling is given
by 
\begin{equation}
V(\delta k)=\nu\sum_{j_{y}}u_{\delta k}[j_{y}]u_{-\delta k}[j_{y}].
\end{equation}
It is easy to show that it can be expanded in terms of even powers
of $\delta k$ and that the leading order is $\nu$ 
\begin{equation}
V(\delta k)=\nu-\sum_{n\ge1}\nu_{n}\delta k^{2n}.\label{eq:effectiveparcoup}
\end{equation}
The edge state is unstable over the finite quasimomentum interval
where $V(\delta k)>|\Delta(\delta)|$. Following the general procedure
presented in Appendix \ref{sec:Generalized-normal-mode}, we find
the eigenenergies and amplification amplitude of the Bogoliubov edge
state normal modes 
\begin{equation}
E(\delta k)=\tilde{E}(\delta k),\quad\lambda(\delta k)=\sqrt{V(\delta k)^{2}-\Delta(\delta k)^{2}}.
\end{equation}
and 
\begin{equation}
E(\delta k)=\tilde{E}(\delta k)+\sqrt{\Delta(\delta k)^{2}-V(\delta k)^{2}}.
\end{equation}
in the unstable and stable quasimomentum ranges, respectively.

We can recover the edge state dispersion and lineshape of the amplification
amplitude obtained numerically and shown in Fig 2 of the main text
by keeping the leading order contributions in Eqs. (\ref{eq:energyedge},\ref{eq:detedge},\ref{eq:effectiveparcoup}),
\begin{eqnarray}
V(\delta k) & \approx & \nu-\nu_{1}\delta k^{2}\,\quad\nu_{1}=0.014,\\
\tilde{E}(\delta k) & \approx & v\delta k,\quad\frac{v}{J}=\frac{{\rm d}\varepsilon}{{\rm d}k_{x}}\Big|_{k_{p}/2}=-1,\\
\Delta(\delta k) & \approx & J\frac{{\rm d}^{2}\varepsilon}{{\rm d}k_{x}^{2}}\Big|_{k_{p}/2}\frac{\delta k^{2}}{2},\quad\frac{{\rm d}^{2}\varepsilon}{{\rm d}k_{x}^{2}}\Big|_{k_{p}/2}=1.4.
\end{eqnarray}

\section{Input-output formalism\label{sec:Input-output-formalism}}

We include the effects of dissipation using the standard input-output
formalism \cite{GerryKnight}. Each site is described by the Langevin
equation 
\begin{equation}
\dot{\hat{a}}_{\mathbf{j}}=i[\hat{H},\hat{a}_{\mathbf{j}}]-\kappa_{\mathbf{j}}\hat{a}_{\mathbf{j}}/2+\sqrt{\kappa_{\mathbf{j}}}\hat{a}_{\mathbf{j}}^{({\rm in)}}.\label{eq:Langevin}
\end{equation}
Here,  $\kappa_{\mathbf{j}}$ is the decay rate on site $\mathbf{j}$.
On the sites coupled to waveguides, the decay rates $\kappa_{{\rm in}}$,
$\kappa_{{\rm out}}$, and $\kappa_{{\rm sink}}$ are induced by the
coupling to the waveguides and are chosen to achieve impedance matching.
In this case, the input fields $\hat{a}_{\mathbf{j}}^{({\rm in)}}$
describe the field impinging on site $\mathbf{j}$ from the corresponding
waveguide including the field vacuum fluctuations. The field $\hat{a}_{\mathbf{j}}^{({\rm out)}}$
leaking out of the waveguide is given by the input-output relations
\begin{equation}
\hat{a}_{\mathbf{j}}^{(out)}=\hat{a}_{\mathbf{j}}^{(in)}-\sqrt{\kappa_{\mathbf{j}}}\hat{a}_{\mathbf{j}}.\label{eq:inputoutput}
\end{equation}
On the remaining sites (i.e. those not coupled to iwaveguides), we
assume a small uniform decay rate $\kappa$ $(\kappa_{\mathbf{j}}=\kappa)$
corresponding to internal loss; $\hat{a}_{\mathbf{j}}^{({\rm in)}}$
describes the corresponding incident vacuum fluctuations associated
with this loss port. 

The linear response and the noise properties of the array are simple
functions of the retarded Green's functions 
\begin{eqnarray}
\tilde{G}_{E}(\omega,\mathbf{j},\mathbf{j'}) & = & -i\int_{-\infty}^{\infty}dt\Theta(t)e^{i\omega t}\langle[\hat{a}_{\mathbf{j}}(t),\hat{a}_{\mathbf{j'}}^{\dagger}(0)]\rangle,\\
\tilde{G}_{I}(\omega,\mathbf{\mathbf{j}},\mathbf{j'}) & = & -i\int_{-\infty}^{\infty}dt\Theta(t)e^{i\omega t}\langle[\hat{a}_{\mathbf{j}}^{\dagger}(t),\hat{a}_{\mathbf{j'}}^{\dagger}(0)]\rangle.
\end{eqnarray}
We calculate the above Green's functions numerically for a finite
array.

\subsection*{Response Ellipses}

In Fig. \ref{fig:Linear-response}a (Fig. \ref{fig:Linear-response}b)
we show the linear response of the pair of quadratures
\begin{eqnarray}
\hat{X}_{\mathbf{j}} & = & \frac{1}{\sqrt{2}}\left(e^{i\theta_{\mathbf{j}}/2}\hat{a}_{\mathbf{j}}^{\dagger}(t)+e^{-i\theta_{\mathbf{j}}/2}\hat{a}_{\mathbf{j}}(t)\right),\nonumber \\
\hat{Y}_{\mathbf{j}} & = & \frac{i}{\sqrt{2}}\left(e^{i\theta_{\mathbf{j}}/2}\hat{a}_{\mathbf{j}}^{\dagger}(t)-e^{-i\theta_{\mathbf{j}}/2}\hat{a}_{\mathbf{j}}(t)\right).\label{eq:quaddef}
\end{eqnarray}
to a classical field of frequency $\omega_{p}/2$ injected at the
input (output) port as a function of the field phase $\vartheta$,
\[
\langle\hat{a}_{\mathbf{j'}}^{(in)}\rangle=\sqrt{\kappa_{\mathbf{j'}}}e^{i\vartheta},
\]
$\mathbf{j}'$ indicates the site attached to the relevant injection
waveguide ($\mathbf{j}'=\mathbf{j}_{{\rm in}}$ for Fig. \ref{fig:Linear-response}a,
$\mathbf{j}'=\mathbf{j}_{{\rm out}}$ for Fig. \ref{fig:Linear-response}b).
The response on each site is represented by the ellipse
\[
\begin{pmatrix}\langle\hat{X}_{\mathbf{j}}\rangle\\
\langle\hat{Y}_{\mathbf{j}}\rangle
\end{pmatrix}=\begin{pmatrix}\cos\gamma_{\mathbf{jj'}} & -\sin\gamma_{\mathbf{jj'}}\\
\sin\gamma_{\mathbf{jj'}} & \cos\gamma_{\mathbf{jj'}}
\end{pmatrix}\begin{pmatrix}r_{\mathbf{jj'}}^{+}\cos(\vartheta-\eta_{\mathbf{jj'}})\\
r_{\mathbf{jj'}}^{-}\sin(\vartheta-\eta_{\mathbf{jj'}})
\end{pmatrix},
\]
where 
\[
r_{\mathbf{jj'}}^{\pm}=\kappa_{\mathbf{j}'}\left(|\tilde{G}_{E}(0,\mathbf{j},\mathbf{j}')|\pm|\tilde{G}_{I}(0,\mathbf{j},\mathbf{j}')|\right)
\]
are the major and minor semiaxes of the ellipse, and 
\begin{equation}
\gamma_{\mathbf{jj'}}=\arg\left(\frac{G_{E}(0,\mathbf{j},\mathbf{j}^{'})}{G_{I}(0,\mathbf{j},\mathbf{j}^{'})}\right)\Big/2-\theta_{\mathbf{j}}/2\label{eq:gamma}
\end{equation}
is the angle parametrizing the orientation of the major semiaxis.
In the figure, the reference angle $\gamma_{\mathbf{j}\mathbf{j'}}=0$
is rotated by $\pi/2$ compared to the page vertical. The maximum
(minimum) response $r_{\mathbf{jj'}}^{+}$ ($r_{\mathbf{jj'}}^{-}$)
at site $\mathbf{j}$ to an input field at the port $\mathbf{j}'$
is obtained for the driving phase $\vartheta=\eta_{{\rm \mathbf{jj'}}}$
($\vartheta=\eta_{{\rm \mathbf{jj'}}}+\pi/2$ ),
\begin{equation}
\eta_{{\rm \mathbf{jj'}}}=\pi/2-\arg\left(G_{E}(0,\mathbf{j},\mathbf{j}')G_{I}(0,\mathbf{j},\mathbf{j}^{'})\right)\big/2.\label{eq:eta}
\end{equation}

\subsection*{Gain, Reverse Gain, and Input Reflection }

As we analyze our amplifier in the phase-sensitive mode of operation,
it is useful to consider its scattering properties in a quadrature
representation, using an optimal quadrature basis fields in the input
and output waveguides. The relevant scattering between these waveguides
is described by
\begin{eqnarray*}
\hat{{\cal X}}_{{\rm out}}[\omega] & = & s_{{\cal X},{\cal U}}[\omega]\hat{{\cal U}}_{{\rm in}}[\omega]+s_{{\cal X},{\cal V}}[\omega]\hat{{\cal V}}_{{\rm in}}[\omega]+...\\
\hat{{\cal Y}}_{{\rm out}}[\omega] & = & s_{{\cal Y},{\cal U}}[\omega]\hat{{\cal U}}_{{\rm in}}[\omega]+s_{{\cal Y},{\cal V}}[\omega]\hat{{\cal V}}_{{\rm in}}[\omega]+...
\end{eqnarray*}
where we have omitted writing terms describing contributions from
vacuum noise incident from the internal loss ports, as well as terms
describing the reflection of noise incident from the output waveguide
(due to slightly imperfect impedance matching). We define the quadratures
of the output waveguide as 
\begin{eqnarray*}
\hat{{\cal X}}_{{\rm out}}(t) & = & \frac{1}{\sqrt{2}}\left(e^{i\theta_{{\rm out}}}\hat{a}_{\mathbf{j_{{\rm out}}}}^{({\rm out)}\dagger}(t)+e^{-i\theta_{{\rm out}}}\hat{a}_{\mathbf{j_{{\rm out}}}}^{({\rm out)}}(t)\right),\\
\hat{{\cal Y}}_{{\rm out}}(t) & = & \frac{i}{\sqrt{2}}\left(e^{i\theta_{{\rm out}}}\hat{a}_{\mathbf{j_{{\rm out}}}}^{({\rm out)}\dagger}(t)-e^{-i\theta_{{\rm out}}}\hat{a}_{\mathbf{j_{{\rm out}}}}^{({\rm out)}}(t)\right).
\end{eqnarray*}
while the quadratures of the input-waveguide are defined as
\begin{eqnarray*}
\hat{{\cal U}}_{{\rm in}}(t) & = & \frac{1}{\sqrt{2}}\left(e^{i\theta_{{\rm in}}}\hat{a}_{\mathbf{j_{{\rm in}}}}^{({\rm in)}\dagger}(t)+e^{-i\theta_{{\rm in}}}\hat{a}_{\mathbf{j_{{\rm in}}}}^{({\rm in)}}(t)\right),\\
\hat{{\cal V}}_{\mathbf{{\rm in}}}(t) & = & \frac{i}{\sqrt{2}}\left(e^{i\theta_{{\rm in}}}\hat{a}_{\mathbf{j_{{\rm in}}}}^{({\rm in)}\dagger}(t)-e^{-i\theta_{{\rm in}}}\hat{a}_{\mathbf{j_{{\rm in}}}}^{({\rm in)}}(t)\right).
\end{eqnarray*}
We will define both the quadrature basis in each coupling waveguide
so that $|s_{{\cal X},{\cal U}}[\omega=0]|$ is maximal. This implies
that the largest amplification of a narrow-band signal centered at
zero frequency ($\omega_{p}/2$ in the lab frame) occurs when the
incident signal is in the $\hat{{\cal U}}_{{\rm in}}$ quadrature
of the input waveguide, with the amplified output being contained
in the output field $\hat{{\cal X}}_{{\rm out}}$ quadrature of the
output waveguide. It also follows naturally that for vacuum noise
inputs, $\hat{{\cal Y}}_{{\rm out}}$ will be the optimally squeezed
quadrature. An explicit calculation of the scattering matrix in terms
of the Green's functions introduced above shows that the angles defining
these preferred input and output quadratures are given by 

\begin{equation}
\theta_{{\rm in}}=\eta_{\mathbf{j}_{{\rm out}}\mathbf{j}_{{\rm in}}},\quad\theta_{{\rm out}}=\gamma_{\mathbf{j}_{{\rm out}}\mathbf{j}_{{\rm in}}}+\theta_{\mathbf{j_{{\rm out}}}}/2,\label{eq:thetainthetaout}
\end{equation}
see Eqs. (\ref{eq:gamma},\ref{eq:eta}). The fact that in general
$\theta_{{\rm in}}\neq\theta_{{\rm out}}$ is a simple reflection
of the fact that our system can both perform phase-sensitive amplification
of incident signals, as well as simply rotate incident signals in
phase space. 

The frequency dependent power gain is then by definition given by
the transmission probability:

\begin{eqnarray*}
G(\omega) & = & |s_{{\cal X},{\cal U}}(\omega)|^{2}\\
 & = & \kappa_{{\rm in}}\kappa_{{\rm out}}\Big|e^{i(\theta_{{\rm in}}-\theta_{{\rm out}})}G_{E}(\omega,\mathbf{j}^{({\rm out)}},\mathbf{j}^{({\rm in})})\\
 &  & +e^{i(\theta_{{\rm in}}+\theta_{{\rm out}})}G_{I}(\omega,\mathbf{j}^{({\rm out)}},\mathbf{j}^{({\rm in})})\\
 &  & +e^{i(\theta_{{\rm out}}-\theta_{{\rm in}})}G_{E}^{*}(-\omega,\mathbf{j}^{({\rm out)}},\mathbf{j}^{({\rm in})})\\
 &  & +e^{-i(\theta_{{\rm in}}+\theta_{{\rm out}})}G_{I}^{*}(-\omega,\mathbf{j}^{({\rm out)}},\mathbf{j}^{({\rm in})})\Big|^{2}\Big/4
\end{eqnarray*}
The reverse gain is obtained similarily by exchanging the two indices
of the Green's functions in the above formula and in Eq. (\ref{eq:thetainthetaout}).
Finally, the input reflection probability describes the reflection
of signals incident in the (amplified) ${\cal U}$ quadrature of the
input port 
\begin{eqnarray*}
R[\omega] & = & |s_{{\cal U},{\cal U}}(\omega)|^{2}\\
 &  & \Big(\Big|e^{i\theta_{{\rm in}}}\left(1-i\kappa_{{\rm in}}G_{E}(\omega,\mathbf{j}^{({\rm out)}},\mathbf{j}^{({\rm in})})\right)\\
 &  & +e^{-i\theta_{{\rm in}}}\left(1+i\kappa_{{\rm in}}G_{I}^{*}(-\omega,\mathbf{j}^{({\rm out)}},\mathbf{j}^{({\rm in})})\right)\Big|\\
 &  & +\Big|e^{i\theta_{{\rm in}}}\left(1-i\kappa_{{\rm in}}G_{I}(\omega,\mathbf{j}^{({\rm out)}},\mathbf{j}^{({\rm in})})\right)\\
 &  & +e^{-i\theta_{{\rm in}}}\left(1+i\kappa_{{\rm in}}G_{E}^{*}(-\omega,\mathbf{j}^{({\rm out)}},\mathbf{j}^{({\rm in})})\right)\Big|\Big)^{2}\Big/4.
\end{eqnarray*}

\subsection*{Added noise and squeezing}

The symmetrized frequency-resolved noise in the amplified and de-amplified
output-waveguide output field quadratures shown in Figure \ref{fig:Quantum-stationary-state}(a)
are defined as 
\begin{eqnarray*}
S_{{\cal X},{\cal X}}(\omega) & = & \int_{-\infty}^{\infty}\frac{dt}{2}e^{i\omega t}\Big\{\hat{{\cal X}}_{{\rm out}}(t),\hat{{\cal X}}_{{\rm out}}(0)\Big\},\\
S_{{\cal Y},{\cal Y}}(\omega) & = & \int_{-\infty}^{\infty}\frac{dt}{2}e^{i\omega t}\Big\{\hat{{\cal Y}}_{{\rm out}}(t),\hat{{\cal Y}}_{{\rm out}}(0)\Big\},
\end{eqnarray*}
respectively. The number of added noise quanta shown in Figure \ref{fig:Quantum-stationary-state}(b)
is defined by,
\[
S_{{\rm add}}(\omega)\equiv\frac{S_{{\cal X},{\cal X}}(\omega)}{G(\omega)}-\frac{1}{2}.
\]

\subsection*{Noise Ellipses}

In Figure \ref{fig:Quantum-stationary-state}(c), the stationary state
of each site $\mathbf{j}$ is represented as a noise ellipse. The
noise ellipse on a particular site $\mathbf{j}$ is obtained by diagonalizing
the corresponding covariance matrix 
\begin{eqnarray*}
V_{\mathbf{j}} & = & \begin{pmatrix}\langle\hat{X}_{\mathbf{j}}^{2}\rangle & \langle\hat{X}_{\mathbf{j}}\hat{Y}_{\mathbf{j}}+\hat{Y}_{\mathbf{j}}\hat{X}_{\mathbf{j}}\rangle\\
\langle\hat{X}_{\mathbf{j}}\hat{Y}_{\mathbf{j}}+\hat{Y}_{\mathbf{j}}\hat{X}_{\mathbf{j}}\rangle & \langle\hat{Y}_{\mathbf{j}}^{2}\rangle
\end{pmatrix}\\
 & = & \begin{pmatrix}\cos\tilde{\gamma}_{\mathbf{j}} & \sin\tilde{\gamma}_{\mathbf{j}}\\
-\sin\tilde{\gamma}_{\mathbf{j}} & \cos\tilde{\gamma}_{\mathbf{j}}
\end{pmatrix}\begin{pmatrix}\left(\tilde{r}_{\mathbf{j}}^{+}\right)^{2} & 0\\
0 & \left(\tilde{r}_{\mathbf{j}}^{-}\right)^{2}
\end{pmatrix}\begin{pmatrix}\cos\tilde{\gamma}_{\mathbf{j}} & -\sin\tilde{\gamma}_{\mathbf{j}}\\
\sin\tilde{\gamma}_{\mathbf{j}} & \cos\tilde{\gamma}_{\mathbf{j}}
\end{pmatrix},
\end{eqnarray*}
and identifying the square root of its eigenvalues $\tilde{r}_{\mathbf{j}}^{+}$
and $\tilde{r}_{\mathbf{j}}^{-}$ as the ellipse semi-axes and the
angle $\tilde{\gamma}_{\mathbf{j}}$ as the ellipse rotation angle
(the reference angle $\tilde{\gamma}_{\mathbf{j}}=0$ is rotated by
$\pi/2$ compared to the page vertical).

\section{Effective model for the edge state coupled to waveguides\label{sec:Appeffectivemodel}}

In this supplementary note we want to derive an effective quantum
field theory for the edge state coupled to a waveguide. We want to
model a finite system which will then support a single edge state
with a ring geometry (i.e.~periodic boundary conditions). We adopt
the simplest possible approach valid when the input signal has a bandwidth
well within the amplifier bandwidth. In this case, we can approximate
the edge state velocity and amplification rates to be constant, $\tilde{E}\approx v\delta k$,
$\Delta\approx0$, and $V(\delta k)\approx\nu$, respectively. Moreover,
we can neglect the quasimomentum dependence of the edge state transverse
wavefunction. We thus arrive at the edge state ladder operator 
\begin{equation}
\hat{c}(j_{\parallel})=\sqrt{\frac{1}{N}}\sum_{n=1}^{N}e^{i\delta k_{n}j_{\parallel}}\hat{\alpha}_{\delta k}\approx e^{-i\kappa_{p}j_{\parallel}/2}\sum_{j_{\perp}}u[j_{\perp}]\hat{a}_{j_{\parallel},j_{\perp}},\label{eq:edgepositionresolveddiscrete}
\end{equation}
where $N$ is the number of sites along the edge, and the indexes
$j_{\parallel}$ and $j_{\perp}$ ($j_{\perp}\ge1$) label the position
in the directions parallel and longitudinal to the edges (for the
edge state along the upper edge described above $j_{\parallel}=j_{x}$
and $j_{\perp}=-j_{y}$). Moreover, $u(j_{\perp})$ is the transverse
edge state wavefunction, $u(j_{\perp})\equiv u_{k_{p}/2}(-j_{\perp})$
see Eqs. (\ref{eq:matrixeom1},\ref{eq:polenergy}). The periodic
boundary conditions $\hat{c}(j_{\parallel})=\hat{c}(N+j_{\parallel})$
are enforced by the quasimomentum quantization $\delta k_{n}=2\pi n/N$.

We consider a waveguide attached to a single site along the edge in
the logitudinal position $j_{\parallel}\equiv j_{{\rm ex}}$. The
coupling to the waveguide as described by standard input-output theory
is entirely characterized by the decay rate $\kappa$. Since, the
edge state is the only state within the amplifier bandwidth the coupling
of the waveguide to a single site is equivalent to a direct coupling
to the edge state at the same longitudinal position $j_{{\rm ex}}$
with the renormalized decay rate $\kappa'=\kappa|u[j_{\perp}=1]|^{2}$.
The resulting Langevin equation reads 
\begin{align}
\dot{\hat{c}}(j_{\parallel})=i[\hat{H}^{({\rm edge})},\hat{c}(j_{\parallel})]-\frac{\kappa'}{2}\delta_{j_{\parallel},j_{{\rm ex}}}\hat{c}(j_{\parallel})+\sqrt{\kappa'}\delta_{j_{\parallel},j_{{\rm ex}}}\hat{a}^{({\rm in})},\label{eq:cavity_eom}
\end{align}
with the input-output relations 
\begin{equation}
\hat{a}^{({\rm out})}=\hat{a}^{({\rm in})}-\sqrt{\kappa'}\hat{c}(j_{{\rm ex}})
\end{equation}

Next we take the continoum limit by defining the chiral edge field
\begin{equation}
\hat{c}(z)=\sqrt{\frac{1}{L}}\sum_{n=-\infty}^{\infty}e^{i\delta k_{n}j_{\parallel}}\hat{\alpha}_{\delta k},\label{eq:edgepositionresolvedcont}
\end{equation}
where $L$ is the edge length and the periodic boundary conditions
$\hat{c}(z)=\hat{c}(z+L)$ follow from the quasimomentum quantization
$\delta k_{n}=2\pi n/L$. From Eq.~(\ref{eq:Hamedge}) with $\tilde{E}=v\delta k$,
$\Delta=0$, and $V(\delta k)=\nu$, and Eq. (\ref{eq:cavity_eom})
we find the field equations 
\begin{equation}
(\partial_{t}+v\partial_{z})\hat{c}(z)=\nu\hat{c}^{\dagger}(z)-\frac{\kappa'}{2}\delta(z-z_{{\rm ex}})\hat{c}(z)+\sqrt{\kappa'}\delta(z-z_{{\rm ex}})\hat{a}^{({\rm in})}.\label{eq:Heisembergtoy}
\end{equation}
We note that due to the point-interaction with the waveguide the field
$\hat{c}(z)$ is not continous at the position $z=z_{{\rm ex}}$ where
the waveguide is attached, $\hat{c}(z_{{\rm ex}}+0_{+})\neq\hat{c}(z_{{\rm ex}}+0_{-})$
($0_{+}$ and $0_{-}$ are infinitesimal positive and negative numbers,
respectively). Keeping this in mind, the input output relation in
the continous limit reads 
\begin{equation}
\hat{a}^{({\rm out})}=\hat{a}^{({\rm in})}-\sqrt{\kappa'}\left(\hat{c}(0_{+})+\hat{c}(0_{-})\right)/2.\label{eq:outputinputcont}
\end{equation}

Before investigating the interaction with the waveguide, we first
discuss the propagation inside the ring. For concreteness we consider
$v>0$ such that $\hat{c}(z_{{\rm ex}}+0_{+})$ ($\hat{c}(z_{{\rm ex}}+0_{-})$)
is the field immediately after (before) the interaction with the waveguide.
A signal travels an almost complete round trip from $z=z_{{\rm ex}}+0_{+}$
to $z=L+z_{{\rm ex}}+0_{+}=z_{{\rm ex}}+0_{-}$ in the time $t=L/v$.
During this time no interaction with the waveguide takes place. From
Eq. (\ref{eq:Heisembergtoy}), one readily finds
\begin{eqnarray*}
\hat{X}(z_{{\rm ex}}+0_{-},L/v+t_{0}) & = & e^{\nu L/v}\hat{X}(z_{{\rm ex}}+0_{+},t_{0}),\\
\hat{Y}(z_{{\rm ex}}+0_{-},L/v+t_{0}) & = & e^{-\nu L/|v|}\hat{Y}(z_{{\rm ex}}+0_{+},t_{0}).
\end{eqnarray*}
Here, we have introduced the amplified and de-amplified quadratures,
$\hat{X}=(\hat{c}^{\dagger}+\hat{c})/\sqrt{2}$ and $\hat{Y}=i(\hat{c}^{\dagger}-\hat{c})/\sqrt{2}.$
We can conclude that a signal with the right phase during a complete
round trip inside the ring experience a power gain 
\begin{equation}
G=e^{2\nu L/v}.\label{eq:gain}
\end{equation}
Next, we discuss the interaction with the waveguide. An input signal
from the waveguide will be partly reflected and partly transmitted
into the ring at $z=0_{+}$. Then, it will propagate inside the ring
until it has completed a round trip. A signal with the right phase
will be amplified along the way. Before starting a new round trip,
part of the amplified signal returns into the waveguide. If the signal
remaining in the ring at the beginning of the second round trip is
smaller compared to the signal in the ring at the beginning of the
first round trip, the signal will decay after few round trips. In
this regime, the waveguide stabilizes the edge state. By integrating
the Heisenberg equations (\ref{eq:Heisembergtoy}) close to $z=0$
we find 
\begin{eqnarray}
\begin{pmatrix}\sqrt{|v|}\hat{c}(0_{+})\\
\hat{a}^{({\rm out})}
\end{pmatrix}=\begin{pmatrix}r & t\\
-t & r
\end{pmatrix}\begin{pmatrix}\sqrt{|v|}\hat{c}(0_{-})\\
\hat{a}^{({\rm in})}
\end{pmatrix}\label{eq:inoutcont}
\end{eqnarray}
where $r$ and $t$ are the reflection and transmission probability
amplitudes, 
\begin{equation}
r=\frac{4-g^{2}}{4+g^{2}},\qquad t=\frac{4g}{4+g^{2}},\quad g=\sqrt{\frac{\kappa'}{v}}.\label{eq:inoutcont2}
\end{equation}
To prevent instability, we require that the transmission $t$ be large
enough that the field in the edge-mode ring does not grow with each
round trip. The simplest case is where we tune the decay rate $\kappa$
so that all of the incident wave in the edge mode ends up in the coupling
waveguide, i.e. $r=0$, $t=1$. This requires $g=2$, or in terms
of the decay rate $\kappa$, the edge state velocity $v$, and the
edge state transversal wavefunction at the edge $u(j_{\perp}=1)$,
we find 
\begin{equation}
\kappa=\frac{4v}{|u(j_{\perp}=1)|^{2}}.\label{eq:impmatchcont}
\end{equation}
If this impedance matching condition is met, signals incident from
the waveguide in the $\hat{X}$ quadrature will be reflected back
with a power gain $G=G_{{\rm Max}}$ independent of frequency. The
impedance matching ensures that multiple traversals of the ring are
impossible, which both precludes instability, but also prevents the
formation of standing wave resonances and a strongly frequency-dependent
gain.

We note that the effective model described above can be straightforwardly
extended to describe a chiral edge state coupled to several waveguides.
In particular, we consider a set up with two impedance matched waveguides
at position $z=z_{{\rm in}}$ and $z=z_{{\rm out}}$ ($z_{{\rm out}}-z_{{\rm in}}=L$),
respectively. A signal from the first waveguide entering the edge
state at position $z=z_{{\rm in}}$ is entirely transmitted into the
second waveguide at position $z=z_{{\rm out}}$. For the appropriate
phase of the input signal, we have 
\begin{equation}
\hat{{\cal U}}^{{\rm in}}(t)=\sqrt{v}\hat{X}(z_{{\rm in}}+0_{+},t),\label{eq:inedgeampl}
\end{equation}
where $\hat{{\cal U}}^{({\rm in})}$ is the input quadrature. We denote
$\hat{{\cal X}}^{{\rm out}}$, the corresponding amplified quadrature
at $z=z_{{\rm out}}$, 
\begin{equation}
\hat{{\cal X}}^{{\rm out}}(t)=-\sqrt{v}\hat{X}(z_{{\rm out}}+0_{-},t).\label{eq:edgeoutampl}
\end{equation}
From Eq. (\ref{eq:Heisembergtoy}) one finds 
\begin{equation}
|\hat{{\cal X}}^{{\rm out}}(t+L/v)|^{2}=G|\hat{{\cal U}}^{{\rm out}}(t)|^{2}
\end{equation}
where the gain $G$ is given by Eq. (\ref{eq:gain}).

\subsection*{Comparison between finite size simulations and effective model}

In the regime where our effective theory applies (for signals well
within the amplifier bandwidth), it can reproduce well the simulations
of finite arrays. There are some quantitative differences due to well
understood finite size effects. For instance, the edge state velocity
is not constant in the finite size array but rather decreases close
to the edges where the edge state propagation changes direction. For
this reason the gain does not depend only on the number of sites separating
the input and the output port but also on the precise position of
the ports. We have placed the input and output port close to the edges
to enhance the gain. Indeed, the gain in the finite size simulations
is slightly larger than predicted by Eq.~(\ref{eq:gain}). Due to
the position dependent edge state velocity also the value of the decay
rate required to obtain impedance matching depends on the position
where the waveguide is attached (and is lower at the edges).

We note that the analytic continuum model can be extended to capture
the frequency dependence of the gain and noise of our travelling wave
amplifier; one needs however to incorporate into the model the leading-order
quasimomentum dependence of the edge state velocity (which creates
an effective pump detuning for the relevant parametric process). This
is will be presented in a future work.

\section{Quantitative analysis of the resilience of the topological amplifier
to losses\label{sec:Applosses}}

Here, we show that our quantum amplifier design offer some degree
of resilience to intrinsic losses and the corresponding noise. We
want to calculate noise which is added to a signal propagating from
the input to the output port when intrinsic losses are present. In
addition to the impedance matched outcoupling towards the waveguides
at the input and output ports, we consider a loss channel on each
site and denote the corresponding decay rate as $\kappa_{\mathbf{j}}^{({\rm loss})}(z)$.

In the continous limit, and with the same approximations as above,
we find the quantum field equations valid in the region between the
input and output port
\begin{eqnarray}
(\partial_{t}+v\partial_{z})\hat{X}(z,t) & = & (\nu-\kappa^{({\rm loss})}(z)/2)\hat{X}(z,t)\nonumber \\
 &  & \sqrt{\kappa^{({\rm loss})}(z)}\hat{X}^{({\rm loss})}(z,t).\label{eq:added_noise_eom}
\end{eqnarray}
 where $\hat{X}^{({\rm loss})}(z,t)$ is the vacuum noise: $\langle\hat{X}^{({\rm loss})}(z,t)\rangle=0$
and 
\begin{equation}
\langle\hat{X}^{({\rm loss})}(z,t)\hat{X}^{({\rm loss})}(z',t')\rangle=\frac{1}{2}\delta(z-z')\delta(t-t').\label{eq:noisecorrelators}
\end{equation}
We fourier transform into frequency space and find the solution
\begin{eqnarray}
\hat{X}[z,\omega] & = & e^{[(i\omega+\nu)(z-z_{{\rm in}})-\int_{z_{{\rm in}}}^{z}\kappa^{({\rm loss})}(z')dz'/2]/v}\hat{X}(z_{{\rm in}}+0_{+},\omega)\nonumber \\
 &  & +\frac{1}{v}\int_{z_{{\rm in}}}^{z}dz'e^{[(i\omega+\nu)(z-z')-\int_{z'}^{z}\kappa^{({\rm loss})}(z'')dz''/2]/v}\nonumber \\
 &  & \times\sqrt{\kappa^{({\rm loss})}(z')}\hat{X}^{({\rm loss})}(z',\omega).\label{eq:freqspacquadwithnoise}
\end{eqnarray}
 Putting together Eq.~(\ref{eq:freqspacquadwithnoise}) for $z=z_{{\rm out}}-0_{-}$
with Eqs.~(\ref{eq:inedgeampl},\ref{eq:edgeoutampl}) we can relate
the output signal $\hat{{\cal X}}^{{\rm out}}$ to the input signal
$\hat{{\cal U}}^{{\rm in}}$ of the amplifier
\begin{eqnarray}
\hat{{\cal X}}^{{\rm out}}[\omega] & = & -e^{[(i\omega+\nu)L-\int_{z_{{\rm in}}}^{z_{{\rm out}}}\kappa^{({\rm loss})}(z')dz'/2]/v}\hat{{\cal U}}^{({\rm in})}[\omega]\nonumber \\
 & + & \frac{1}{\sqrt{v}}\int_{z_{{\rm in}}}^{z_{{\rm out}}}dz'e^{[(i\omega+\nu)({z_{{\rm out}}}-z')-\int_{z'}^{z_{{\rm out}}}\kappa^{({\rm loss})}(z'')dz''/2]/v}\nonumber \\
 &  & \times\sqrt{\kappa^{({\rm loss})}(z')}\hat{X}^{({\rm loss})}(z',\omega).\label{eq:xout}
\end{eqnarray}
 The added noise $S_{{\rm add}}(\omega)$ in noise quanta is by definition,
\begin{equation}
S_{{\rm add}}(\omega)=\frac{S_{{\cal X,X}}(\omega)}{G(\omega)}-\frac{1}{2}.\label{eq:Sadd}
\end{equation}
where $S_{{\cal X,X}}(\omega)$ is the symmetrized noise at the output
port, 
\begin{equation}
S_{{\cal X,X}}[\omega]\delta(\omega+\omega')=\frac{1}{2}\langle\{\hat{{\cal X}}[\omega],\hat{{\cal X}}[\omega']\}\rangle.
\end{equation}
and $G(\omega)$ is the gain. We note that in our simple approach
$S_{{\cal X,X}}$, $G$ and $S_{{\rm add}}$ are frequency independent.
From Eqs.~(\ref{eq:xout},\ref{eq:noisecorrelators}) we find 
\begin{equation}
G=e^{[2\nu L-\int_{z_{{\rm in}}}^{z_{{\rm out}}}\kappa^{({\rm loss})}(z')dz']/v}
\end{equation}
and 
\begin{equation}
S_{{\rm add}}=\frac{1}{v}\int_{0}^{L}dre^{-[2\nu r-\int_{0}^{L}\kappa^{({\rm loss})}(r+z_{{\rm in}})dr']/v}\kappa^{({\rm loss})}(r+z_{{\rm in}}).
\end{equation}
From this expression we see that for $2\nu>\kappa^{({\rm loss})}(z)$
the noise added at position $r+z_{{\rm in}}$ is cut off exponentially
as a function of the distance $r$ from the input. In particular,
in the large gain limit and assuming that $\kappa^{({\rm loss})}(z)$
is smooth close to $z=z_{{\rm in}}$, we can approximate 
\begin{equation}
S_{{\rm add}}\approx\frac{\kappa^{({\rm loss})}(z_{{\rm in}})}{2\nu}.
\end{equation}
Note crucially that as one increases the length $L$ of the amplifying
channel, the gain increases exponentially, while the added noise remains
constant. We thus see explicitly that the amplifying channel is immune
to the majority the internal loss noise in the system.

\end{document}